\documentclass[12pt]{article}
\usepackage{amsmath}
\usepackage{amssymb}
\usepackage{graphicx,bbm,mathrsfs}
\usepackage{color}

\newcommand{\beq}{\begin{eqnarray}} 
\newcommand{\eeq}{\end{eqnarray}} 
\newcommand{\tb}{\tan\beta} 
 
\usepackage{listings}
\usepackage{caption}
\usepackage{cite}

\newcommand{\bea}{\begin{eqnarray}}  
\newcommand{\eea}{\end{eqnarray}}
\newcommand{\nbea}{\begin{align*}}
\newcommand{\neea}{\end{align*}}
\newcommand{\nbeq}{\begin{equation*}}
\newcommand{\neeq}{\end{equation*}}
\newcommand{\bear}{\begin{eqnarray}}  
\newcommand{\eear}{\end{eqnarray}}

 \newcommand{\comment}[1]{}
\usepackage{array}
\newcolumntype{M}[1]{>{\centering\arraybackslash}m{#1}}
\newcolumntype{N}{@{}m{0pt}@{}}

\numberwithin{equation}{section}

\begin{document}

\begin{flushright}\small{KCL-PH-TH/2016-42, CERN-TH/2016-164~~~~~~~~~} \end{flushright} 

\vspace*{5mm} 

\begin{center}
 
\mbox{\large\bf Doubling Up on Supersymmetry in the Higgs Sector} \\
\vspace{2mm}

\vspace*{8mm}

{\sc John Ellis$^{1,2}$}, {\sc J\'er\'emie Quevillon$^1$ and {\sc Ver\'onica Sanz$^3$}} 

\vspace{6mm}

{\small 
$^1$ Theoretical Particle Physics \& Cosmology Group, Department of Physics, \\ 
King's College London,  Strand, London WC2R 2LS, United Kingdom \\
\vspace{2mm}
$^2$ Theoretical Physics Department, CERN, CH 1211 Geneva 23, Switzerland \\
\vspace{2mm}
$^3$ Department of Physics and Astronomy, University of Sussex, Brighton BN1 9QH, UK \\}
\end{center}

\vspace*{10mm}

\begin{abstract}
\vspace*{5mm}

We explore the possibility that physics at the TeV scale possesses approximate
$N = 2$ supersymmetry, which is reduced to the $N=1$ minimal supersymmetric
extension of the Standard Model (MSSM) at the electroweak scale. This doubling
of supersymmetry modifies the Higgs sector of the theory, with consequences for
the masses, mixings and couplings of the MSSM Higgs bosons, whose
phenomenological consequences we explore in this paper. The mass of the lightest
neutral Higgs boson $h$ is independent of $\tan \beta$ at the tree level, and the 
decoupling limit is realized whatever the values of the heavy Higgs boson masses. Radiative
corrections to the top quark and stop squarks dominate over those due to particles
in $N=2$ gauge multiplets. 
We assume that these radiative corrections fix $m_h \simeq 125$~GeV,
whatever the masses of the other neutral Higgs bosons $H, A$,
a scenario that we term the $h$2MSSM.
Since the $H, A$ bosons decouple from the $W$ and $Z$ bosons
in the $h$2MSSM at tree level, only the LHC constraints on $H, A$ and $H^\pm$ couplings to
fermions are applicable. These and the indirect constraints from LHC measurements of $h$ couplings are consistent with
$m_A \gtrsim 200$~GeV for $\tan \beta \in (2, 8)$ in the $h$2MSSM.

\end{abstract}
\vspace*{1.5cm}
\begin{flushleft}\small{July 2016} \end{flushleft} 

\newpage

\section{Introduction}

Since the Standard Model is chiral, it can accommodate only $N=1$ supersymmetry,
as in the minimal supersymmetric extension of the Standard Model (MSSM).
On the other hand, any new physics beyond the Standard Model would contain
vector-like representations of the SU(2)$\times$U(1) gauge group of the Standard Model.
As such, it could accommodate $N=2$ supersymmetry. One could even argue that it
{\it should} possess the maximum possible degree of supersymmetry, namely $N=2$.
Indeed, there are plenty of theoretical set-ups that lead naturally to a chiral $N=1$
supersymmetry model at the electroweak scale with a vector-like $N=2$ extension
at the TeV scale, including models invoking extra dimensions and superstring model
constructions~\cite{delAguila:1984qs,Polonsky:2000zt,Antoniadis:2006uj,N2fromextra}.

Studies of possible $N=2$ extensions of the Standard Model have a long history,
with considerable attention paid to the gauge and matter sectors of such models.
An $N=2$ vector multiplet would contain more degrees of freedom than in the MSSM. 
In particular, gauginos would no longer be Majorana particles, but Dirac. Moreover, 
additional adjoint scalar fields would appear, namely a new singlet $S$, triplet $T$ and octet $O$. 
The phenomenology of the Dirac gauginos has been explored in a number of papers~\cite{dirac-gauginos,muless},
and attention also been paid to the Higgs sector of an $N=2$ extension of the
Standard Model, which has interesting differences from the Higgs sector of the MSSM~\cite{Antoniadis:2006uj}. 
This is a natural entry point into phenomenological studies of $N=2$ models,
since the Higgs sector of the MSSM is necessarily vector-like, and hence readily
modified to realize $N=2$ supersymmetry. Moreover, the exploration of Higgs phenomenology
is well underway, with important experimental
constraints coming from measurements of the $h(125)$ Higgs boson~\cite{h125} and searches
for the heavier MSSM Higgs bosons.

As has been pointed out in previous studies, the $N=2$ version of the tree-level
supersymmetric Higgs potential (\ref{pot}) contains an extra term 
$\frac{1}{2}(g_1^2+g_2^{2})|H_1H_2|^2$, which has important phenomenological
consequences~\cite{Antoniadis:2006uj}. In particular, the masses of the Higgs bosons are independent
of $\tan \beta$ at the tree level, and the rotation from the doublet basis $H_1$, $H_2$ to 
the mass eigenstate basis $h$, $H$ is trivial, so that at the tree level the $N=2$ model realizes
automatically the decoupling limit of the MSSM. Hence the tree-level couplings of the lighter neutral scalar
Higgs boson $h$ are necessarily identical to those of a Standard Model Higgs,
and the heavier neutral scalar boson $H$ plays no role in electroweak
symmetry breaking.

These observations are modified by the radiative corrections to the Higgs sector,
of which the most important are those due to the top-stop sector, as in the MSSM~\footnote{There 
are also tree-level corrections due to the $N=2$ gauge sectors of the theory, 
but it was found in~\cite{Benakli:2012cy} that the contributions of the
additional adjoints $S$ and $T$ to the Higgs boson masses
are typically two orders of magnitude smaller than the loop contributions we consider
below, for values of $m_{S,T} \sim m_{\tilde t}$.}. As in
the MSSM, a practical way to analyze Higgs phenomenology in the model with
$N=2$ supersymmetry is to use the measured mass of the observed Standard Model-like
Higgs boson $m_h \simeq 125$~GeV as a constraint on the other parameters of the model.
In the MSSM case, this has been called the $h$MSSM scenario: the analogous scenario
we propose here is termed the $h$2MSSM scenario.

As we show, an important difference between the $h$MSSM and $h$2MSSM scenarios is that 
the latter can be realized with smaller stop masses than the former for any value of $m_A \gtrsim m_h$,
and for smaller $m_A$ for any fixed values of the stop masses and $\tan \beta$.
This observation then raises the question how light the heavier Higgs bosons $H, A$ can be in the $h$2MSSM,
for what range of $\tan \beta$. 

The LHC constraints on $H \to W^+ W^-$, $Z^0 Z^0$ and $A \to Z h$ decays
are not relevant for the $h$2MSSM, since it realizes automatically the decoupling limit at the tree level,
and the $H W^+ W^-$ and $H Z^0 Z^0$ couplings induced at the loop level are relatively small.
On the other hand, LHC constraints on decays of the heavy Higgs bosons into fermions are in principle relevant.
Specifically, the constraint from the search for $H^\pm \to \tau^\pm \nu$ decays is the same
as in the $h$MSSM. Before saying the same for the LHC constraint on $A/H \to \tau^+ \tau^-$,
one must check the near-degeneracy of the $H$ and $A$, as assumed in the experimental analyses.
As we show, in the $h$2MSSM $m_H - m_A$ is actually typically significantly smaller in magnitude than in the $h$MSSM. 
Consequently, the LHC constraints
on $A/H \to \tau^+ \tau^-$ are directly applicable to the $h$2MSSM.

Also, measurements at LHC Run 1 of the couplings of the $h(125)$ to fermions
impose important indirect constraints on the $h$2MSSM in the $(m_A, \tan \beta)$ plane, though they are weaker than
in the $h$MSSM. As we show,
the principal constraints are those on the ratios of $h$ couplings to up-type quarks, down-type quarks and massive vector bosons, and that on the $h\gamma \gamma$ coupling. We find that the direct searches for heavy Higgs bosons exclude ranges of $m_A$
when $\tan \beta \gtrsim 7$, and the $h$ coupling measurements require $m_A \gtrsim 185$~GeV in the $h$2MSSM, compared with
$\gtrsim 350$~GeV in the $h$MSSM.

This paper is organized as follows. In Section~\ref{sec:setup}, we show the differences between the MSSM
and the $N=2$ Higgs sector, at the tree level in Section~\ref{sec:tree} and including radiative corrections in Section~\ref{subsec:rad},
and we use the dominant loop corrections from the stop sector in both the $h$MSSM and the $h$2MSSM
to evaluate possible stop masses in Section~\ref{subsec:stop}. 
Constraints from the LHC are studied in Section~\ref{sec:couplings}, where we discuss the current direct
constraints from searches for $H, A$ and $H^\pm$ in Section~\ref{subsec:HAsearches}, bounds on the $N=2$
Higgs sector from $h f {\bar f}$, $hW^+W^-$ and $hZ^0Z^0$ couplings in Section~\ref{subsec:run1} and those
from the $h \gamma \gamma$ and $hgg$ couplings in Section~\ref{subsec:hgamgam}. We also discuss
the sizes of anomalous couplings of $h(125)$ that could be constrained by future measurements in Section~\ref{subsec:anom}.
We conclude in Sec.~\ref{sec:concl}.

\section{The $N=2$ Supersymmetric Higgs Sector}\label{sec:setup}

\subsection{Model Framework}

The Lagrangian for an $N=2$ extension of the Standard Model possesses an $R$ symmetry,
and its SU(2)$_{R}\times$U(1)$_{R}^{N=2}$-invariant form can be written 
in the $N=1$ language as \cite{delAguila:1984qs,Polonsky:2000zt}:
\begin{eqnarray}
L &=&\frac{1}{8g^2}[W^{\alpha}W_{\alpha}]_F+[\sqrt{2}igY\Phi_{V}X]_F+ h.c.
\nonumber \\
&&
+[2{\text {Tr}}(
\Phi_{V}^{\dagger}{\text{e}}^{2gV}\Phi_{V}{\text{e}}^{-2gV}+X^{\dagger}
{\text{e}}^{2gV}X+Y^{\dagger}{\text{e}}^{-2gV^T}Y)]_D \, ,
\label{L}
\end{eqnarray}
where $\Phi_{V} \equiv \Phi^a_VT^a$ and $V \equiv V^aT^a$, where the
$T^a$ are the gauge group generators.
The second $F$-term in the upper line of (\ref{L}) is the superpotential, whose
only free parameter is the gauge coupling constant $g$. The coupling 
constant of the Yukawa term in the superpotential is 
determined by the gauge coupling due to the SU(2)$_{R}$ global symmetry. 
The SU(2)$_{R}$ symmetry forbids any chiral Yukawa terms,
so that fermion mass generation in the $N=2$ sector is linked to supersymmetry breaking.
We note also that the U(1)$_{R}^{N=2}$ symmetry forbids any mass terms of the form $W_2 \ni \mu^{\prime} XY$,
and specifically that the usual $N=1$ $\mu$ term $W\sim \mu H_1 H_2$ is forbidden by the full $R$-symmetry. A theory with no $\mu$-term would lead to unacceptably light charginos~\cite{randall}, but couplings of the Higgs multiplet to the adjoint scalars of an N=2 gauge sector provide mechanisms to lift the chargino masses and additional $\mu$-like contributions to the scalar potential~\cite{muless}.
Note that, unlike the SU(2)$_{R}$ global symmetry, the $U(1)_{R}^{N=2}$ symmetry can survive supersymmetry breaking.

Finally, the $N=2$ Higgs sector belongs to a hypermultiplet
$\mathbb H=(\mathcal H^c,\mathcal H)$ whose interactions with the
gauge sector are given by the Lagrangian
\begin{equation}
\int d^4\theta \left\{ \mathcal H^\dagger e^{V}\mathcal H+\mathcal H^c
e^{-V}\mathcal H^{c\dagger}\right\}-\left\{\sqrt{2}\,\int d^2\theta
\mathcal H^c \chi\mathcal H+h.c.\right\} \, .
\label{lagH}
\end{equation}
In the following we analyze the phenomenology of this $N=2$ framework for the Higgs sector
of the MSSM.

\subsection{Tree-Level Analysis}\label{sec:tree}

We can write the tree-level $N=2$ Higgs potential in the usual MSSM
notation where $H_{1,2}$ are the lowest components of the chiral
superfields $\mathcal H$ and $\mathcal H^c$ respectively. 
The $H_{2}$ field gives masses to up-type quarks and the $H_{1}$ field gives masses
to down-type fermions. The potential for these
neutral components of the Higgs doublets is
\begin{eqnarray}
V&=&m_1^2 |H_1|^2+m_2^2 |H_2|^2-m_3^2(H_1 H_2+h.c.) \nonumber\\
&+&\frac{1}{8}(g_1^2+g_2^{2})(|H_1|^2-|H_2|^2)^2 +
\frac{1}{2}(g_1^2+g_2^{2})|H_1H_2|^2 \, ,
\label{pot}
\end{eqnarray}
where $m_{i}^2=m_{H_i}^2+\mu^2$ are the effective low-energy mass parameters
including the soft supersymmetry-breaking and $\mu$ terms. In the last line of (\ref{pot}), the first
quartic term is the usual $D$-term of the $N=1$ MSSM, whereas the second is
a specific $N=2$ effect.
This extra quartic term in the potential has interesting
consequences for the minimization of the potential and the Higgs spectrum, as we now review. 

The conditions to have a
vacuum that breaks electroweak symmetry with the correct value of $m_Z$
for a specific value of $\tan \beta$ are:
\begin{eqnarray}
&&\frac{m_Z^2}{2}=-\mu^2+\frac{1}{\tan^2\beta-1}\left(m_{H_1}^2-
m_{H_2}^2\tan^2\beta\right) \, , \label{mZfix}\\ &&
m_A^2=m_{H_1}^2+m_{H_2}^2+2\mu^2+m_Z^2 \, . \label{mAfix}
\end{eqnarray}
We note the difference between (\ref{mAfix}) and the corresponding MSSM minimization
condition $m_A^2=m_{H_1}^2+m_{H_2}^2+2\mu^2$, which has the consequence
that the value of $m_A$ in the $N=2$ model is larger than that in the MSSM for the same input mass parameters. 

In the $(H_1,H_2)$ basis for the two Higgs doublet fields, 
the CP-even $h/H$ mass matrix can be written in terms of the $Z$ and $A$ boson masses and the angle $\beta$.
In the MSSM, the tree-level mass-squared matrix is 
\begin{equation}
\mathcal{M}{^{2,MSSM}_{\text{tree}}}=
\begin{pmatrix}
m_Z^2\cos^2\beta+m_A^2\sin^2\beta & -(m^2_A+m^2_Z)\cos\beta\sin\beta \\
-(m^2_A+m^2_Z)\cos\beta\sin\beta & m_Z^2\sin^2\beta+m_A^2\cos^2\beta
\end{pmatrix} \, .
\label{massmatrixMSSM}
\end{equation}
On the other hand, if the Higgs sector has $N = 2$ supersymmetry, the tree-level mass-squared matrix
is~\cite{Antoniadis:2006uj}:
\begin{equation}
\mathcal{M}{^{2,N2}_{\text{tree}}}=
\begin{pmatrix}
m_Z^2\cos^2\beta+m_A^2\sin^2\beta & - (m^2_A-m^2_Z)\cos\beta\sin\beta \\
- (m^2_A-m^2_Z)\cos\beta\sin\beta & m_Z^2\sin^2\beta+m_A^2\cos^2\beta
\end{pmatrix} \, ,
\label{massmatrixN2}
\end{equation}
where we note the crucial change: $m^2_A+m^2_Z \to m^2_A-m^2_Z$ in the off-diagonal terms from the MSSM case 
(\ref{massmatrixMSSM})~\footnote{We note in passing that there is a missing minus sign in the off-diagonal terms
in Equation (3.12) of~\cite{Antoniadis:2006uj}.}.
The eigenvalues of the matrices (\ref{massmatrixMSSM}, \ref{massmatrixN2})
correspond to the physical masses-squared of the neutral CP-even Higgs bosons. In the MSSM case they are
\begin{equation}
m_{h/H}^{2,MSSM}=\frac{1}{2} \left( m_{A}^{2}+m_{Z}^2 \mp \sqrt{m_{A}^4+m_{Z}^4 -2 m_{A}^2 m_{Z}^2 \cos 4\beta}\right) \, 
\label{MSSMneutral}
\end{equation}
and the mass of the charged Higgs boson is
\begin{eqnarray}
m_{H^\pm}^{MSSM} = \sqrt{ m_A^2 + m_W^2} \, 
\label{eq:MH+}
\end{eqnarray}
at the tree level~\footnote{We note also that the supersymmetric radiative corrections to this relation are known to be small in 
general in this model.}, whereas in the $N=2$ case they are
\begin{equation}
m_{h}^{N2}=m_Z ; \quad m_H^{N2} = m_A  \, ,
\label{N=2neutral}
\end{equation}
and the charged Higgs boson mass is
\begin{equation}
m_{H^\pm}^{N2} = \sqrt{m^2_A+2\,m^2_W} \, .
\label{N=2charged}
\end{equation}
We see that, as in the MSSM, the spectrum of the $N=2$ Higgs sector
is controlled by $m_A$. However, in contrast to the MSSM, it has no dependence on
$\tan\beta$ at the tree level.

\begin{figure}[!]
\begin{center}
\includegraphics[width=0.45\textwidth]{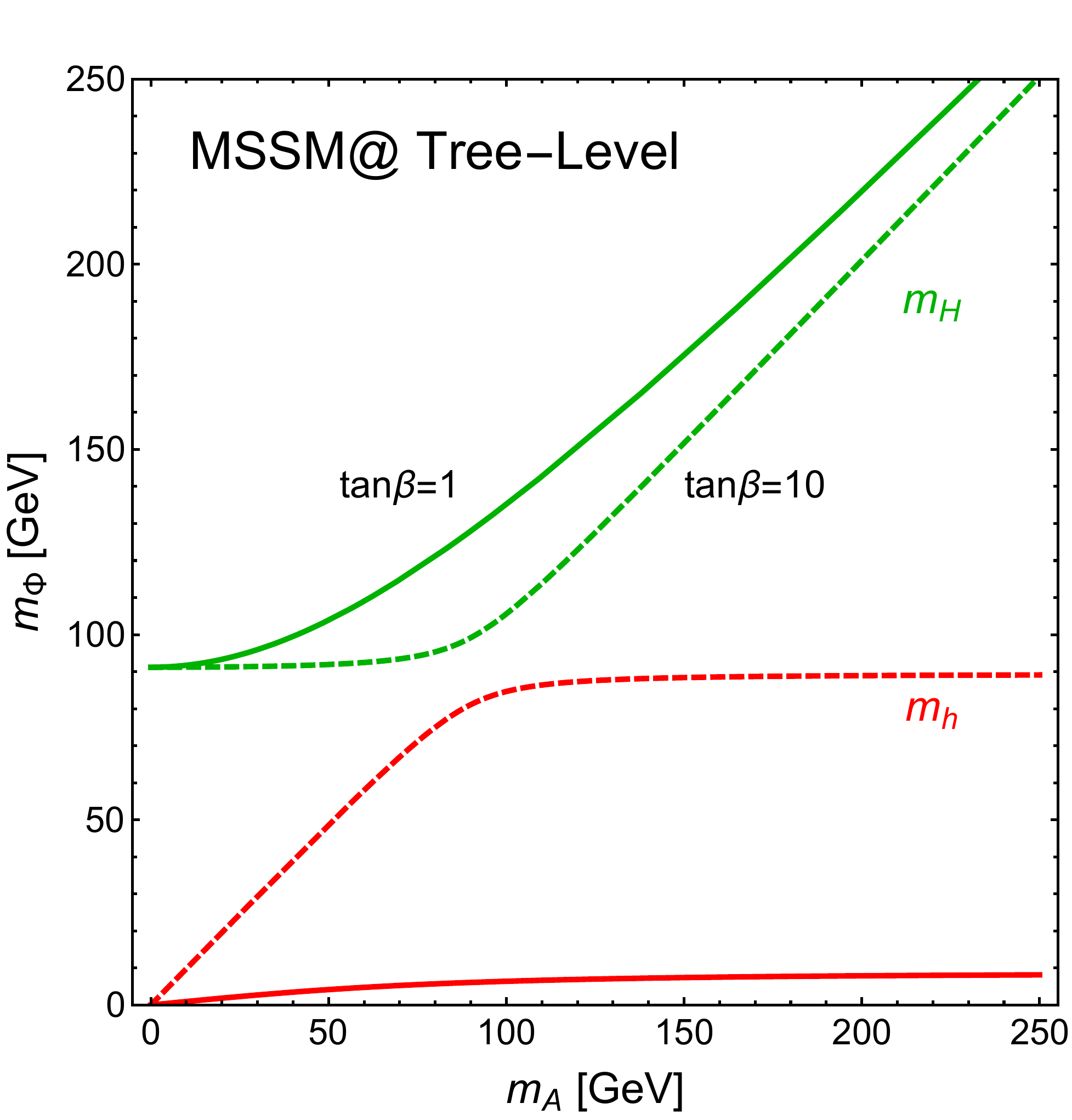} 
\includegraphics[width=0.45\textwidth]{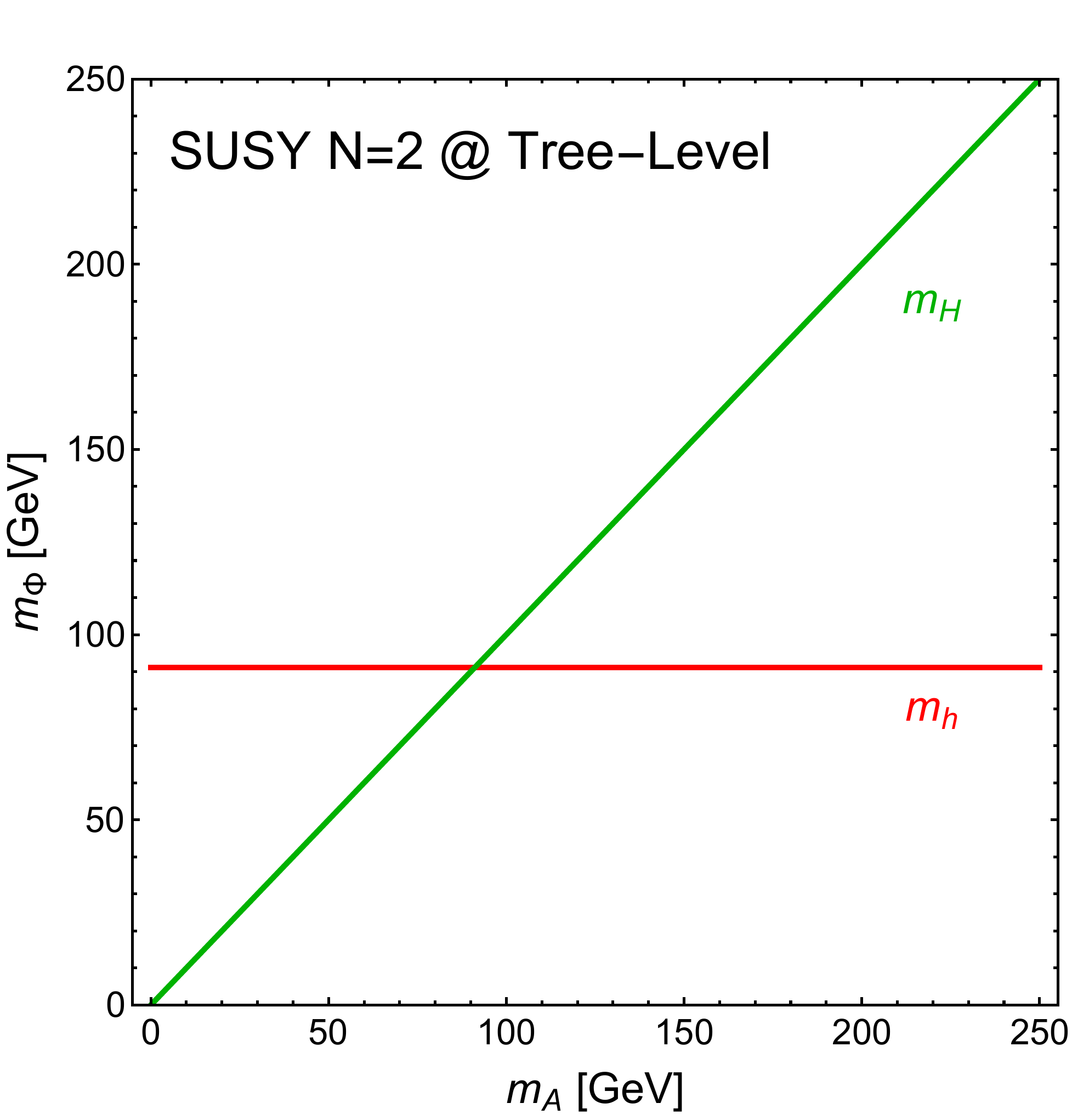} 
\caption{\it The tree-level CP-even Higgs masses $m_h$ (red lines) and $m_H$
(green lines) in the MSSM (left panel, for $\tan \beta = 1$ (solid lines)
and $\tan \beta = 10$ (dashed lines)) and for the $N=2$ MSSM (right panel), as functions of the CP-odd Higgs mass $m_A$.}
\label{masses_tree}
\end{center}
\end{figure}

The left panel of Fig.~\ref{masses_tree} shows the tree-level $N=1$ MSSM CP-even neutral Higgs boson
masses as functions of $m_A$ for different values of $\tan\beta$, 
and we see that $m_h$ increases with $\tan\beta$, its upper limit being $m_Z$.
The right panel of Fig.~\ref{masses_tree} shows the corresponding $N=2$ CP-even neutral Higgs boson masses at the tree level,
where we see that $m_h = m_Z$ independently of $m_A$ and $\tan \beta$, and that $m_H$ crosses $m_h$
without the `level repulsion' effect seen in the left panel.

The physical CP-even Higgs bosons are obtained from the Higgs doublet fields $(H_1, H_2)$
by rotation through an angle $\alpha$:
\begin{equation}
\begin{pmatrix} H \\ h 
\end{pmatrix}
=
\begin{pmatrix}
\cos\alpha & \sin\alpha \\
-\sin\alpha & \cos\alpha
\end{pmatrix}
\begin{pmatrix} H_1 \\ H_2  
\end{pmatrix} \, .
\end{equation}
The MSSM mass-squared matrix (\ref{massmatrixMSSM}) is diagonalized by the following mixing angle:
\begin{equation}
\alpha_{MSSM}=\frac{1}{2} \arctan \left( \tan2\beta \frac{m_{A}^2+m_{Z}^2}{m_{A}^2-m_{Z}^2} \right) \, ,
\end{equation}
which satisfies the relation $-\pi/2 \leq \alpha \leq 0$.
On the other hand, the $N=2$ mass matrix (\ref{massmatrixN2}) is diagonalized by the following mixing angle: 
\begin{equation}
\alpha_{N2}=\beta-\frac{\pi}{2} \, ,
\end{equation}
which also satisfies the relation $-\pi/2 \leq \alpha \leq 0$.

This implies that at the tree level the $N=2$ theory realizes automatically the decoupling limit, in which the lighter CP-even 
neutral Higgs boson $h$ has Standard Model-like couplings and the heavier one, $H$,  does not couple to gauge bosons.

\subsection{Radiative Corrections}\label{subsec:rad}

In our approach, the Higgs sector is described in terms of just the parameters entering the tree-level expressions for 
the masses and mixing, supplemented by the experimentally-known value of $m_h$. In this sense, the $h$MSSM and $h$2MSSM 
approaches can be considered as `model-independent', as the predictions for the properties of the Higgs bosons do not depend 
on the details of the unobserved supersymmetric sector.
We write the mass matrix for the neutral CP-even states as
\begin{equation}
\mathcal{M}^{2}_{\Phi}= {\cal M}^2_{\text{tree}}+
\begin{pmatrix}
\Delta{\cal M}^2_{11} & \Delta{\cal M}^2_{12} \\
\Delta{\cal M}^2_{12} & \Delta{\cal M}^2_{22}
\end{pmatrix} \, ,
\label{massmatrixradcorr}
\end{equation}
where the tree-level matrix $M^{2}_\text{tree}$ is given in (\ref{massmatrixMSSM}) and (\ref{massmatrixN2}) for the MSSM
and its $N=2$ extension, respectively, and the $\Delta{\cal M}^2_{ij}$ are the radiative corrections.

The importance of radiative corrections is manifested by the experimental measurement $m_h = 125$~GeV. 
The most important quantum corrections $\epsilon$ to the CP-even neutral Higgs masses come from
top and stop loops, which alter only the {$\Delta{\cal M}^2_{22}$} element of the mass-squared matrix.
In the MSSM we have:
\begin{equation}
{\mathcal{M}^{2,MSSM}_\Phi}=
\begin{pmatrix}
m_Z^2\cos^2\beta+m_A^2\sin^2\beta & -(m^2_A+m^2_Z)\cos\beta\sin\beta \\
-(m^2_A+m^2_Z)\cos\beta\sin\beta & m_Z^2\sin^2\beta+m_A^2\cos^2\beta +\epsilon_{MSSM} 
\end{pmatrix} \, ,
\label{massmatrixhMSSM}
\end{equation}
where {$\epsilon_{MSSM}$} depends on the top quark mass, the stop masses through the combination
$M_{SUSY} \equiv \sqrt{m_{\tilde{t}_1}m_{\tilde{t}_2}}$, and the mixing parameter in the stop mass matrix, $X_t$. 
A useful approximate expression for $\epsilon_{MSSM}$ is:
\begin{equation}
  \epsilon_{MSSM}=\frac{3 m_t^4}{2\pi^2v^2}\left(\ln\frac{M^{2}_{SUSY}}{m_t^2}+\frac{X_t^2}{2 M^2_{SUSY}}
      \left(1-\frac{X_t^2}{6 M_{SUSY}^2}\right)\right) \, .
      \label{stoptoploop}
\end{equation}
In general MSSM models, the value of $m_h$ is a complicated function on the model parameters,
particularly if one takes into account two- and more-loop effects.

Other radiative corrections to the Higgs mass matrix have been studied in~\cite{hMSSM,LHCHXSWG}. 
Direct analysis of the dominant one-loop 
contributions from top-stop loops shows that the corrections to the {$\Delta{\cal M}^2_{11}$} 
and {$\Delta{\cal M}^2_{12}$} elements of the 
CP-even Higgs mass matrix are proportional to powers of the quantity $\mu X_{t}/M_{SUSY}^2$. 
Consequently, they are negligible to the extent that $\mu X_{t}/M_{SUSY}^2 \lesssim 1$.

In MSSM-like scenarios with $M_{SUSY}$ up to a few TeV, the consideration of the full one-loop 
contributions or of the known two-loop contributions does not alter this simple 
picture~\footnote{For more details about this particular point, the reader should consult references in~\cite{LHCHXSWG}.}. 
When the SUSY scale is very large, additional checks on the value of $m_h$
are required at low $\tan\beta$, for which a comparison with an effective field theory calculation is necessary. 
Results of such an analysis~\cite{Lee} indicate that, even in such heavy-$M_{SUSY}$ scenarios, 
the predictions  of  the $h$MSSM agree within a few percent with the exact results for $m_H$, 
$\alpha$ and $\lambda_{Hhh}$, as long as the condition $\mu X_{t}/M_{SUSY}^2 \lesssim 1$ is satisfied.

For the purposes of our $N=2$ study here, which is restricted to the Higgs sector, 
we follow the philosophy proposed in~\cite{hMSSM,LHCHXSWG}, in which the $h$MSSM scenario was introduced to discuss the
$N=1$ MSSM Higgs sector. The idea is again to use the known output $m_h$ instead of the unknown input $\epsilon$, adjusting
$\epsilon$ so as to obtain $m_{h}=125$ GeV. Here we extend this idea to the $N=2$ case, in a scenario
we call the $h$2MSSM.

In the $N=1$ case, diagonalizing the one-loop corrected mass-squared matrix (\ref{massmatrixhMSSM})
and requiring that one of the eigenvalues of the mass matrix be $m_h = 125$~GeV
yields the following simple analytical formula for $\epsilon$:
\begin{eqnarray}
\epsilon_{MSSM}= \Delta {\cal M}^{2,MSSM}_{22}= \frac{m_{h}^2(m_{A}^2  + m_{Z}^2 -m_{h}^2) - m_{A}^2 m_{Z}^2 
\cos^2 2\beta } { m_{Z}^2 \cos^{2}{\beta}  +m_{A}^2 \sin^{2}{\beta} -m_{h}^2} \, .
\label{dM22MSSM}
\end{eqnarray}
In this $h$MSSM approach the mass of the heavier neutral CP-even $H$ boson and the mixing angle 
$\alpha$ that diagonalises the $h,H$ states are given by the following simple expressions:
\begin{eqnarray}
m_{H}^{2,MSSM} &= &\frac{(m_{A}^2+m_{Z}^2-m_{h}^2)(m_{Z}^2 \cos^{2}{\beta}+m_{A}^2
\sin^{2}{\beta}) - m_{A}^2 m_{Z}^2 \cos^{2}{2\beta} } {m_{Z}^2 \cos^{2}{\beta}+m_{A}^2
\sin^{2}{\beta} - m_{h}^2} \, , \nonumber \\
\ \ \  \alpha_{MSSM} &= &-\arctan\left(\frac{ (m_{Z}^2+m_{A}^2) \cos{\beta} \sin{\beta}} {m_{Z}^2
\cos^{2}{\beta}+m_{A}^2 \sin^{2}{\beta} - m_{h}^2}\right) \, ,
\label{hMSSM-output}
\end{eqnarray}
in terms of the inputs $m_A$, $\tan\beta$ and the mass of the lighter CP-even eigenstate 
$m_h\!=\! 125$ GeV. 

Turning now to the $N=2$ Higgs sector, we can perform the same analysis as before, starting with the mass matrix
\begin{equation}
{\mathcal{M}^{2,N2}_\Phi}=
\begin{pmatrix}
m_Z^2\cos^2\beta+m_A^2\sin^2\beta & (m^2_Z-m^2_A)\cos\beta\sin\beta \\
(m^2_Z-m^2_A)\cos\beta\sin\beta & m_Z^2\sin^2\beta+m_A^2\cos^2\beta +\epsilon_{N2}
\end{pmatrix} \, .
\label{massmatrixh2MSSM}
\end{equation}
Requiring $m_h^{N2}=125$ GeV, we then obtain
\begin{eqnarray}
\epsilon_{N2}= \Delta {\cal M}^{2,N2}_{22}=  \frac{2 (m_{A}^2-m_{h}^2) (m_{h}^2-m_{Z}^2)}{\cos 2 \beta \left(m_{Z}^2-m_{A}^2\right)+m_{A}^2-2 m_{h}^2+m_{Z}^2} \, .
\label{dM22h2MSSM}
\end{eqnarray}
The heavier CP-even mass-squared eigenvalue and the rotation angle of the mass matrix are then found to be
\begin{eqnarray}
m_{H}^{2,N2} &= &m_{A}^2-m_{h}^2+m_{Z}^2 + \frac{2 (m_{A}^2-m_{h}^2) (m_{h}^2-m_{Z}^2)}{\cos 2\beta \left(m_{Z}^2-m_{A}^2\right)+m_{A}^2-2 m_{h}^2+m_{Z}^2} \, , \nonumber \\
\ \ \  \alpha_{N2} &=& - \arctan \left(\frac{\sin2 \beta (m_{A}^2-m_{Z}^2)}{\cos 2\beta \left(m_{Z}^2-m_{A}^2\right)+m_{A}^2-2 m_{h}^2+m_{Z}^2}\right) \, . 
\label{h2MSSM-output}
\end{eqnarray}
We note that in both the $h$MSSM and the $h$2MSSM scenarios there is the same minimal value for $m_A$:
\beq
m_A \; = \; \sqrt{\frac{m_h^2-m_Z^2}{\sin^2\beta}+m_Z^2} \, .
\label{minMA}
\eeq
The general form of the one-loop stop/top contribution to the $\Delta{\cal M}^2_{22}$ element of the CP-even Higgs
mass matrix, $\epsilon_{MSSM}$, is the same as in the $N=1$ MSSM, see (\ref{stoptoploop}), and one can apply the
same arguments about the relative unimportance of other MSSM loop contributions.

However, in the $N=2$ Higgs sector, there are additional loop contributions to the CP-even mass matrix from singlet and triplet
adjoint scalars.
We use the estimate of their contribution from \cite{Benakli:2011kz,Benakli:2012cy},
where more details about the assumptions behind this estimate can be found:
\begin{align}
\frac{32\pi^2}{v^2} \Delta \epsilon_{N2} = & \frac{g_1^2}{2} \ln \frac{m_S^2}{v^2} + \frac{3 g_2^4}{2} \ln \frac{m_T^2}{v^2} \nonumber\\
&+ \frac{g_1^2 g_2^2}{m_S^2 - m_T^2}\bigg[ m_S^2 \ln \frac{m_S^2}{v^2} - m_T^2 \ln \frac{m_T^2}{v^2} - ( m_S^2 - m_T^2) \bigg] \nonumber\\
\overset{ m_S^2 \rightarrow m_T^2}{\longrightarrow}& \frac{1}{2}\bigg( g_1^4+2 g_1^2 g_2^2 + 3 g_2^4 \bigg) \ln \frac{m_T^2}{v^2} \, ,
\label{STbit}
\end{align}
where $m_S,m_T$ are the masses of the adjoint singlet and triplet scalars, respectively.
In the last line of (\ref{STbit}) we show the limiting value when these additional scalars are degenerate in mass. 
In our approximation, the total radiative correction to the mass matrix is then 
$\epsilon_{N2}=\epsilon_{MSSM}+\Delta \epsilon_{N2}$.
The relative orders of magnitude of these two pieces can be estimated from their ratio when the adjoint singlet and triplet are mass degenerate: 
\beq
\frac{\epsilon_{MSSM}}{\Delta \epsilon_{N2}} \simeq 36 \frac{\ln(\frac{M^2_S}{m^2_t})}{\ln(\frac{m^2_T}{v^2})} \, .
\eeq
This shows that $\Delta\tilde{\epsilon}_{N2}$ is relatively unimportant for our current purposes: in our subsequent
numerical estimates we use $m_S = m_T = 1$~TeV as a default.

\begin{figure}[!]
\begin{center}
\includegraphics[width=0.32\textwidth]{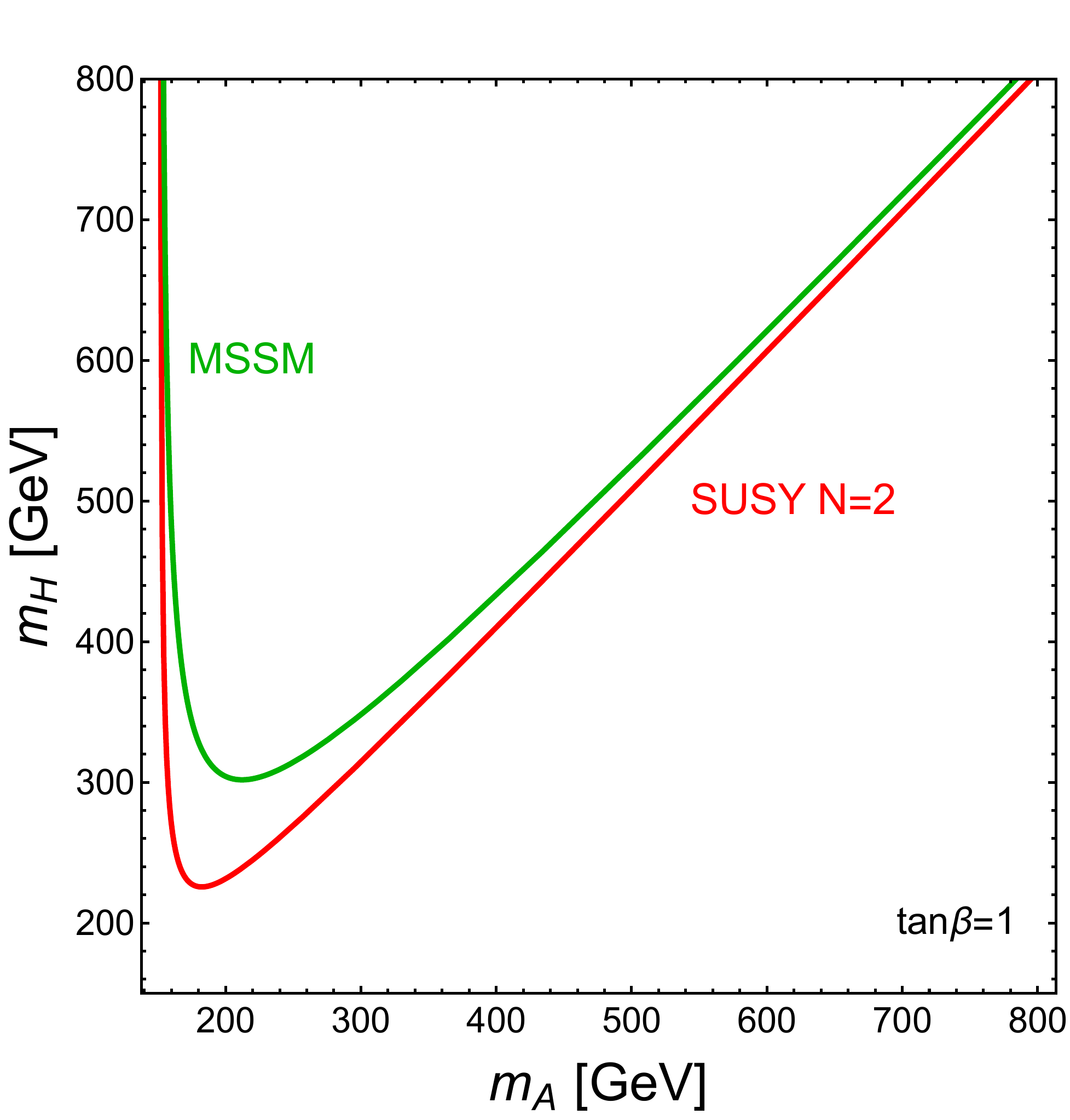} 
\includegraphics[width=0.32\textwidth]{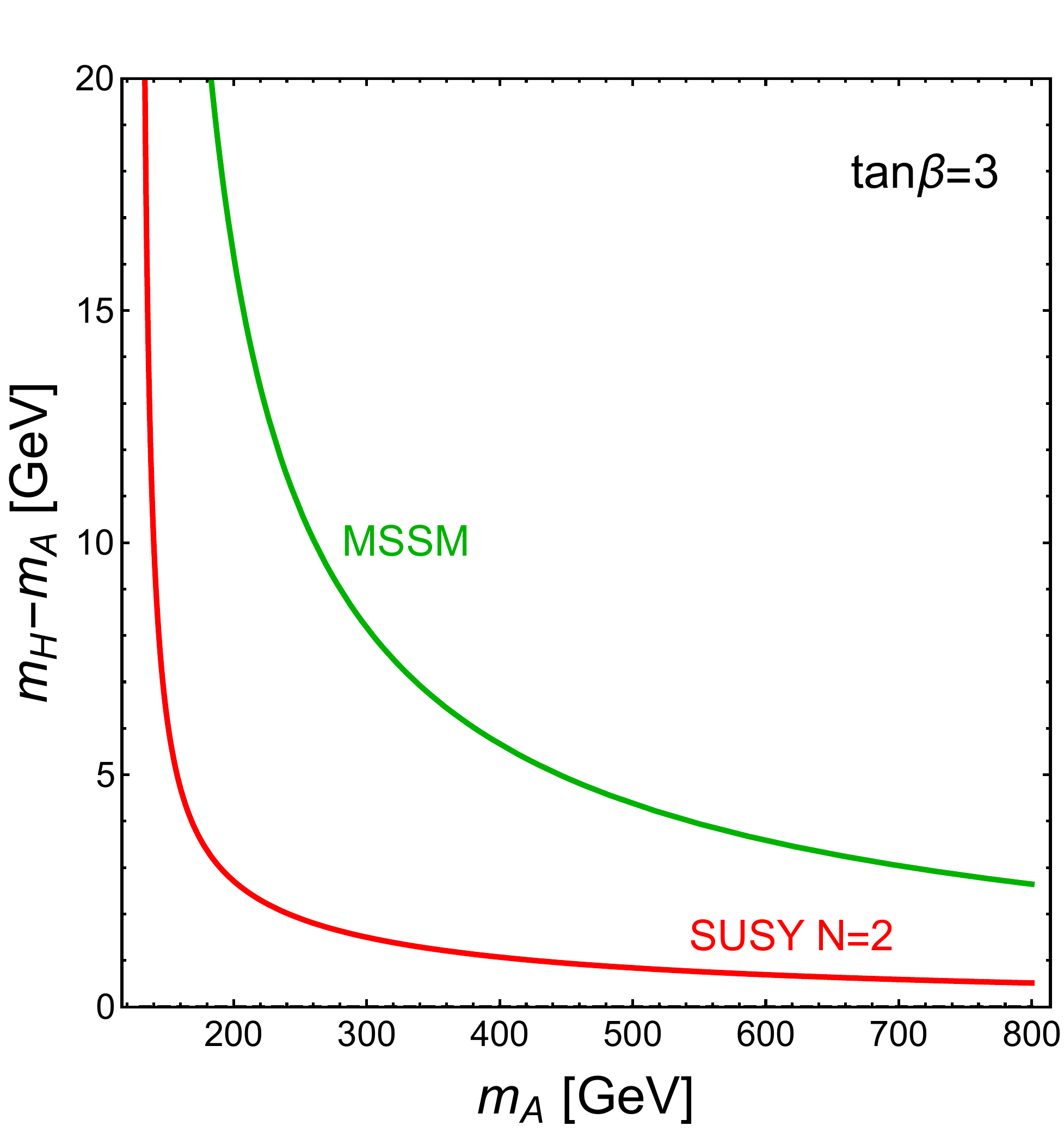}
\includegraphics[width=0.32\textwidth]{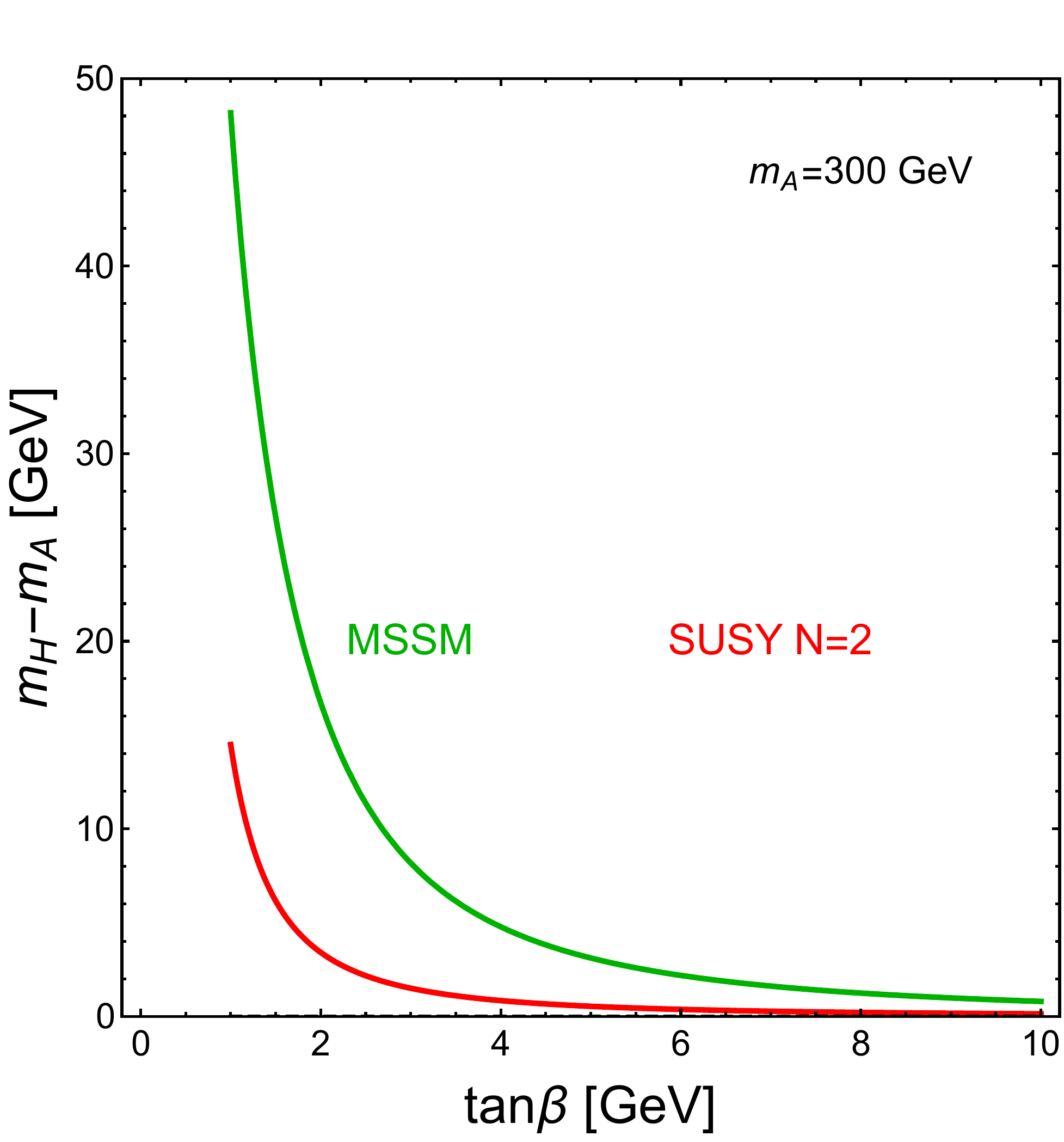} 
\caption{\it Left panel: The values of the mass of the heavier scalar Higgs boson $H$
as functions of $m_A$ for $\tan \beta = 1$, when the leading
one-loop radiative correction to the Higgs mass matrix, $\epsilon$, is chosen such that the 
lighter scalar Higgs boson $h$ has a mass of 125~GeV. Here and in the other panels, the red curve is for the $N=2$ 
$h$2MSSM scenario, and the green curve is for the $N=1$ $h$MSSM.
Middle panel: The mass differences $m_H - m_A$ for $m_h = 125$~GeV in the $N=2$ $h$2MSSM scenario 
and in the $N=1$ $h$MSSM scenario 
as functions of $m_A$ for $\tan \beta = 3$. Right panel: Analogous curves as functions of $\tan \beta$ for $m_A = 300$~GeV.}
\label{masses_1L}
\end{center}
\end{figure}

Fig.~\ref{masses_1L} displays the differences between the $h$MSSM scenario in the $N=1$ case
and the $h$2MSSM scenario in the $N=2$ case.
The left panel of Fig.~\ref{masses_1L} compares the values of the mass of the heavier CP-even Higgs boson $H$
in the $h$2MSSM (red curve) and the $h$MSSM (green curve) as functions of $m_A$ for $\tan \beta = 1$.
We see that the $H$ boson has quite a different mass in the $h$2MSSM as compared to the $h$MSSM.
An interesting point is that, in both scenarios, $m_{H}$ diverges for some specific value of $m_{A}$
slightly above 125 GeV, the exact value depending on $\tan\beta$ as shown in (\ref{minMA}). This corresponds to the fact that 
there is no value of $\epsilon$ that satisfies the requirement $m_{h}=125$~GeV for a region of the
$(m_{A}, \tan\beta)$ parameter plane. 
However, in the $N=2$ $h$2MSSM scenario,
the divergence in the required value of $m_H$ is less severe.

The eagle-eyed reader will notice that the red curve for $m_H$ in the left panel of Fig.~\ref{masses_1L}
lies extremely close to the green curve for $m_A$.
As we see in the other panels of Fig.~\ref{masses_1L}, it is a general feature of the $h$2MSSM that
{$m_H - m_A$ is smaller than in the MSSM. In the middle panel of Fig.~\ref{masses_1L}, we plot the mass splitting 
$m_{H}-m_{A}$ in the $h$2MSSM as a function of $m_{A}$ for $\tan\beta=3$ (red curve).
The right panel of Fig.~\ref{masses_1L} shows the corresponding calculation of the mass splitting 
$m_{H}-m_{A}$ in the $h$2MSSM as a function of $\tan \beta$ for $m_A = 300$~GeV (red curve). The similar
feature of a smaller magnitude is again apparent. The fact that $m_{H}-m_{A}$ is small is
relevant to the LHC experimental searches for $H/A \to \tau^+ \tau^-$ that we discuss later, since they assume that this mass
difference is smaller than their experimental resolution.

Fig.~\ref{cbma2} displays contours of $\cos^2(\beta-\alpha)$ in the $(m_A,\tan\beta)$
plane for the $h$MSSM scenario (left panel) and the $N=2$ $h$2MSSM scenario (right panel).
This quantity determines the coupling of the heavier CP-even Higgs boson $H$ to the electroweak gauge sector. 
We can see that this coupling is significantly reduced in the $h$2MSSM, compared to the $h$MSSM,
reducing the impact of the experimental constraints, as we also discuss later.
\begin{figure}[!]
\begin{center}
\includegraphics[width=0.4\textwidth]{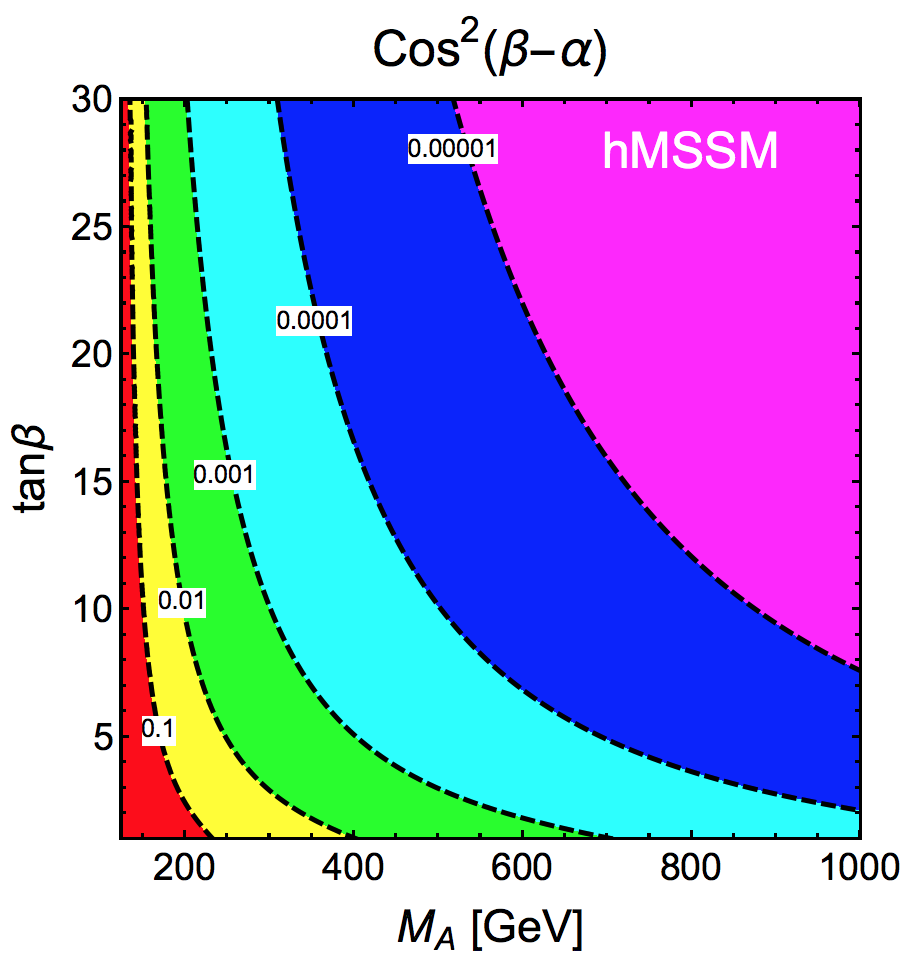} 
\includegraphics[width=0.4\textwidth]{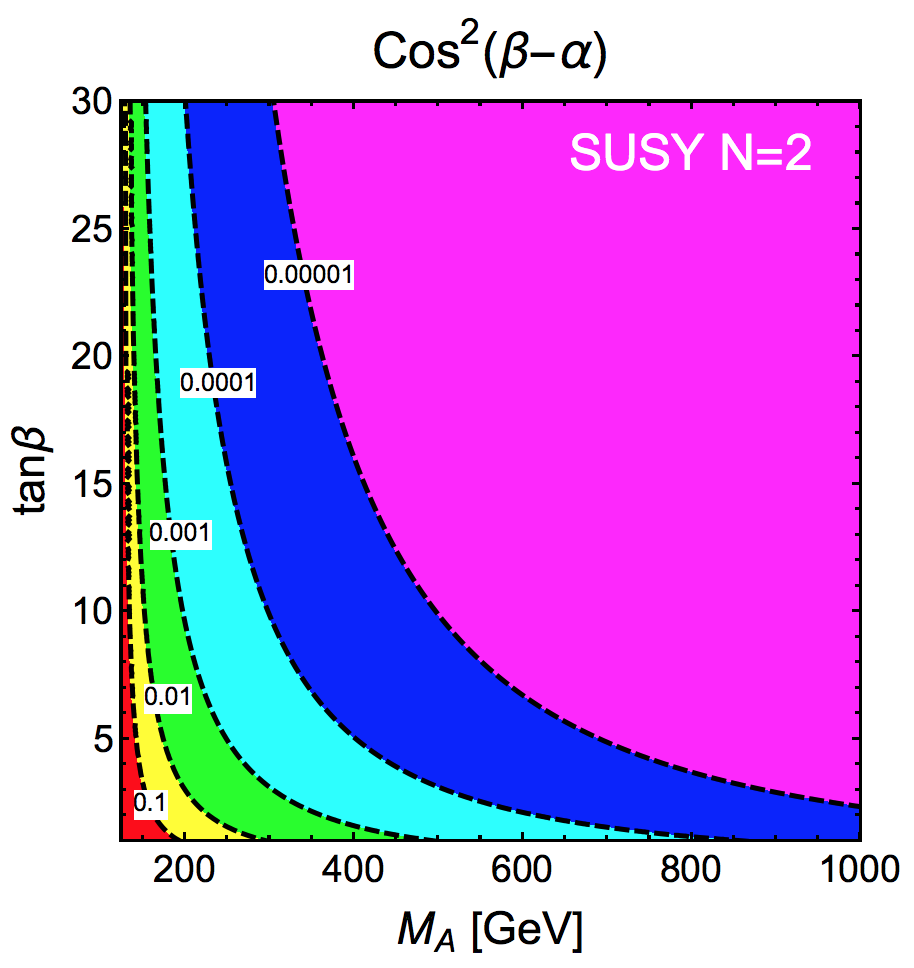} 
\caption{\it Contours of $\cos^2(\beta-\alpha)$ in the $(m_A,\tan\beta)$
plane for the $h$MSSM scenario (left panel) and the $N=2$ $h$2MSSM scenario (right panel).}
\label{cbma2}
\end{center}
\end{figure}
%

\subsection{The Stop Sector in the $h$MSSM and the $h$2MSSM}\label{subsec:stop}

Thus far, we have simply assumed that the stop sector is such that $m_h = 125$~GeV. Now we study
what properties the stop sector must have in order for this to be possible. We recall from (\ref{stoptoploop})
that the two relevant parameters in $\epsilon_{MSSM}$ are $M_{SUSY}$ and $X_t$. As can be seen there, 
the radiative correction increases monotonically with $M_{SUSY}$, but depends in a nontrivial and nonlinear
way on $X_t$. This means that any statement about the required size of $M_{SUSY}$ is dependent on
the assumed value of $X_t$, and more than one value of $X_t$ may yield $m_h = 125$~GeV
with the same value of $M_{SUSY}$. These remarks apply to both the $h$MSSM and the $h$2MSSM.
Looking at Fig.~\ref{masses_tree}, however, we recall that the tree-level value of $m_h$ is larger in the
$N = 2$ extension of the MSSM than in its $N=1$ version. This implies that the required magnitude of 
$\epsilon_{MSSM}$ is smaller in the $h$2MSSM than in the $h$MSSM and hence that, for any fixed
value of $X_t$, the required value of $M_{SUSY}$ is also smaller, as we now discuss in more detail.

We display in Fig.~\ref{XtMSUSY} the values of $M_{SUSY}$ that are required in the $h$MSSM (green dotted lines)
and the $h$2MSSM (red full lines) to yield $m_h = 125$~GeV, as functions of $X_t/M_{SUSY}$. The first point visible in these plots
is that the required value of $M_{SUSY}$ is {\it very} sensitive to $X_t$, in both scenarios. It is occasionally said that
$m_h = 125$~GeV requires, within the MSSM, values of $M_{SUSY}$ in the multi-TeV range. We see that this is true in the 
$h$MSSM for $X_t = 0$ and $\tan \beta = 1$ (left panel), but is {\it not true} in general.
For example, as seen in the middle panel, for most values of $X_t$, $M_{SUSY} < 1000$~GeV is sufficient in the $h$MSSM
if $\tan \beta = 3$, and even $M_{SUSY} < 600$~GeV for a suitable choice of $X_t$. The trend to lower $M_{SUSY}$
continues for $\tan \beta = 10$ (right panel) and larger.

\begin{figure}[!]
\begin{center}
\includegraphics[width=0.32\textwidth]{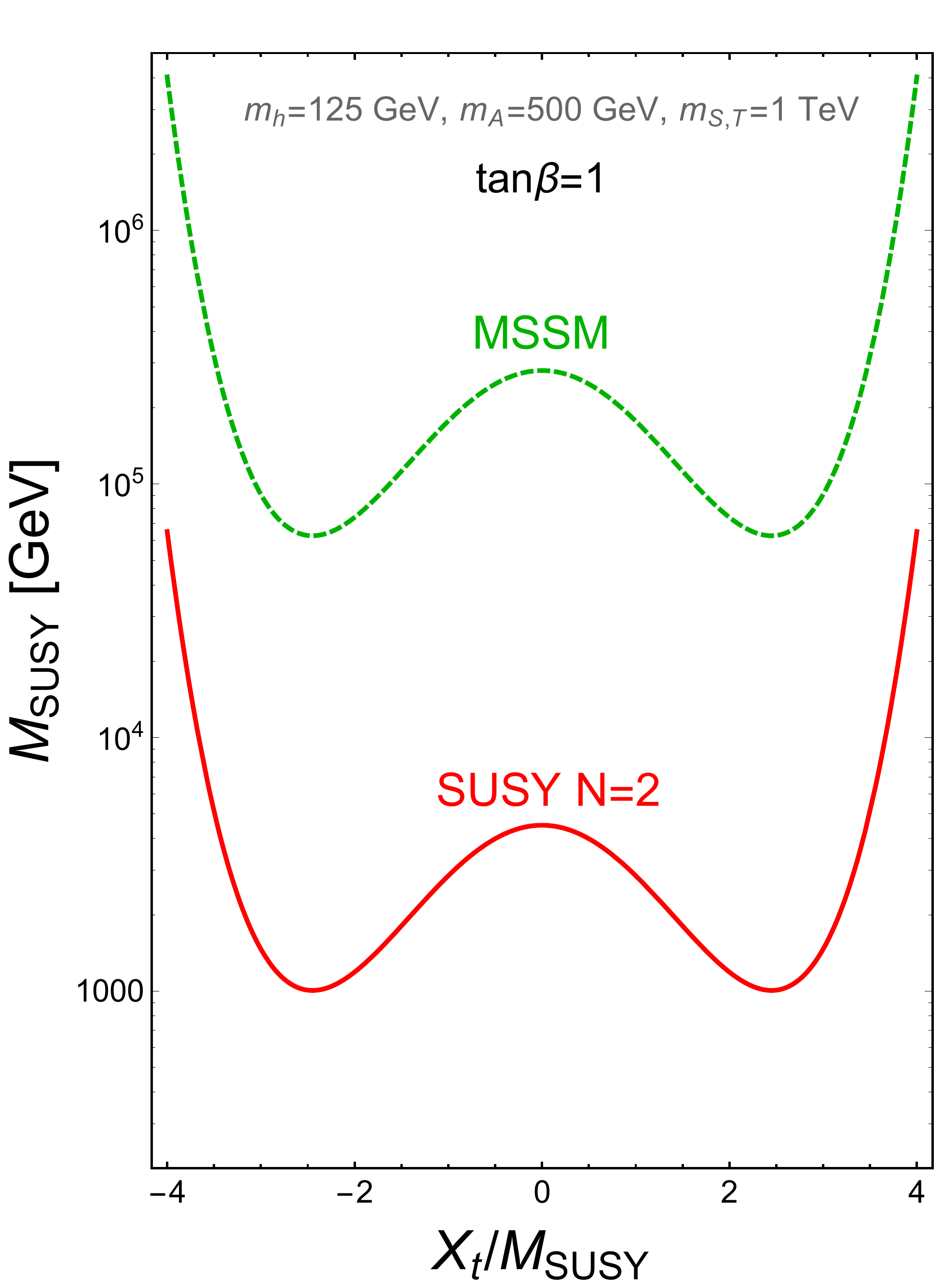} 
\includegraphics[width=0.32\textwidth]{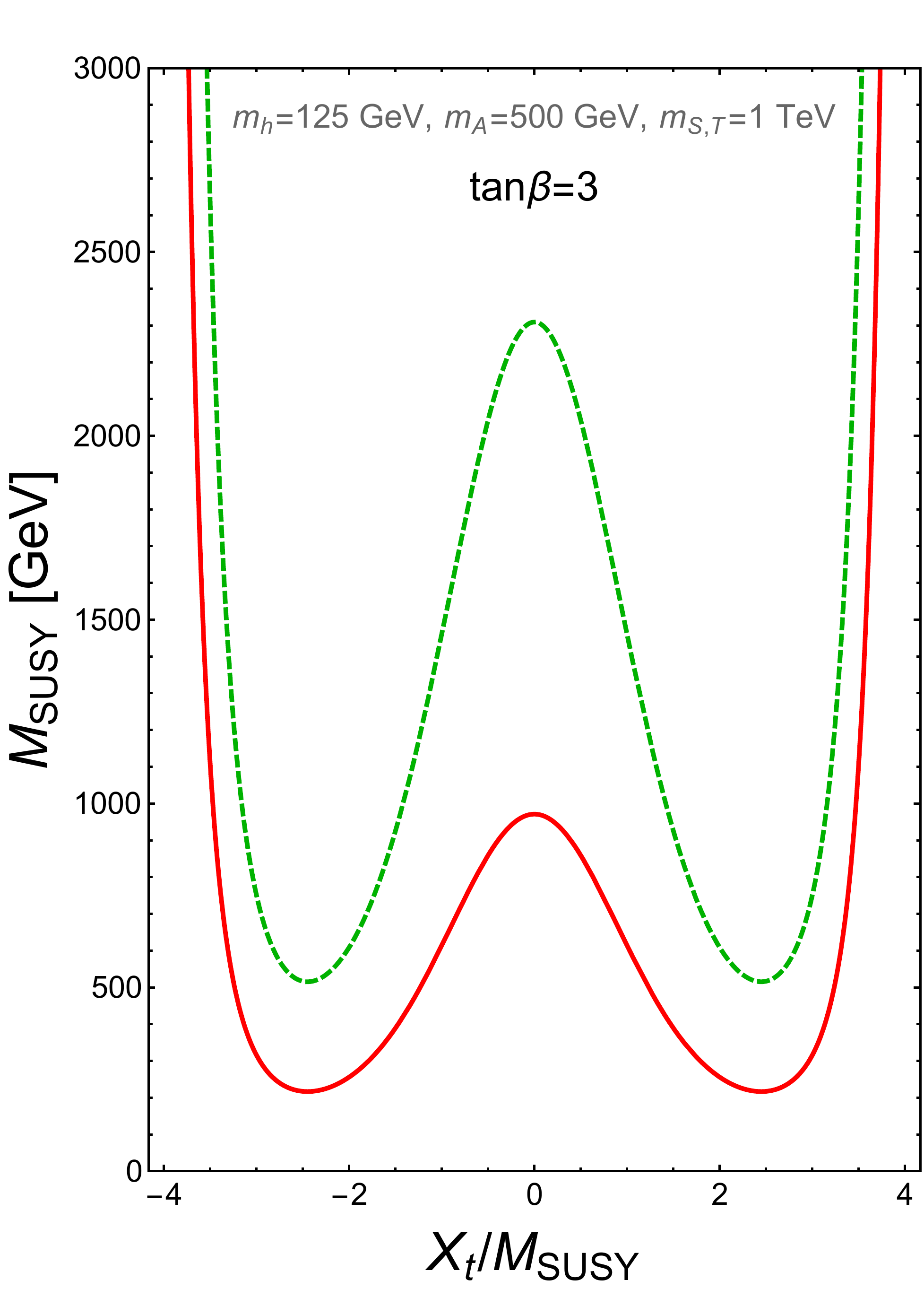} 
\includegraphics[width=0.32\textwidth]{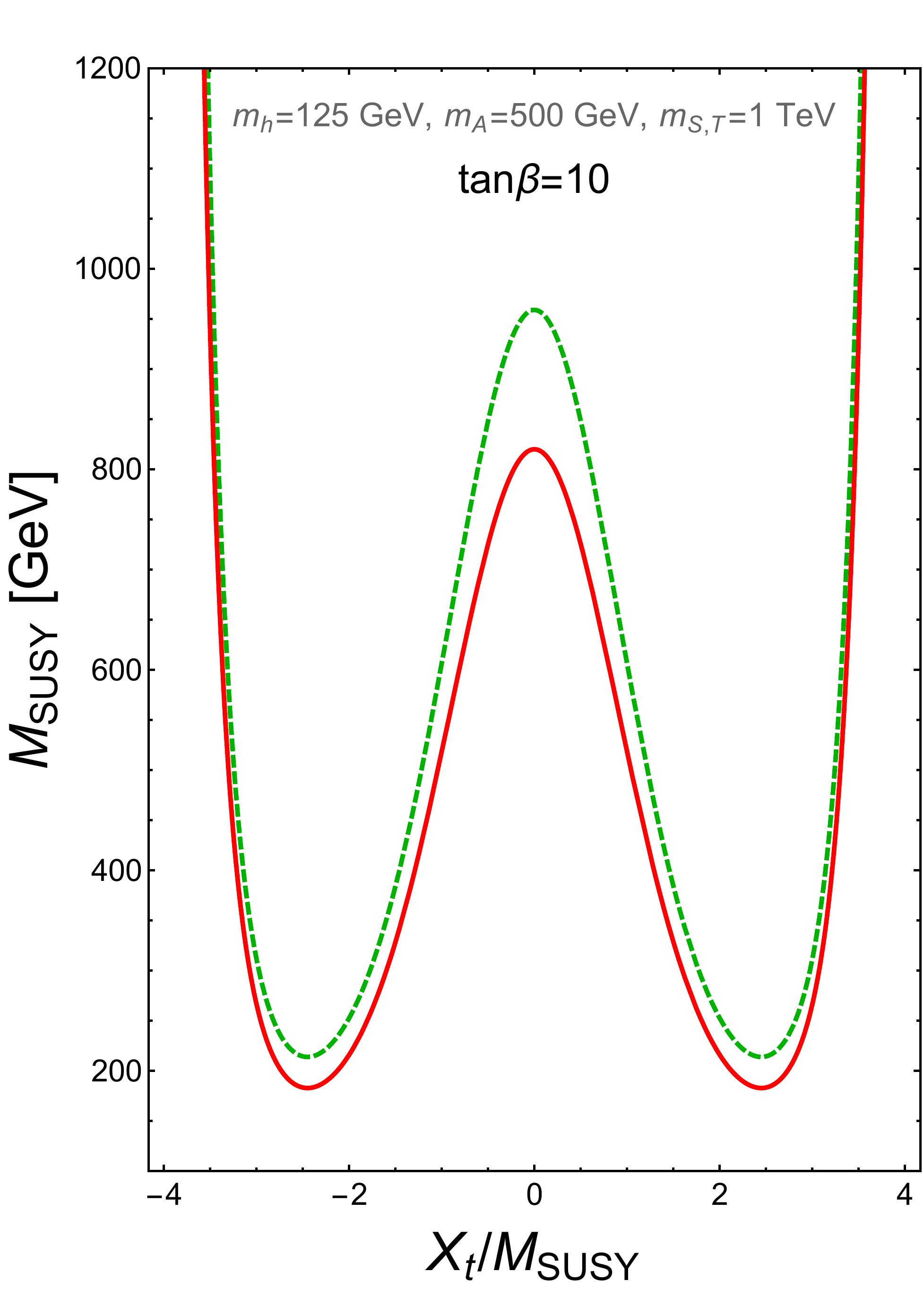} 
\vspace{0.5cm}
\caption{\it Contours of $M_{SUSY}$ as functions of $X_t/M_{SUSY}$ that yield $m_h = 125$~GeV
in the $h$MSSM scenario (green dotted lines) and the $N=2$ $h$2MSSM scenario (red full lines).
The left panel is for $m_A = 500$~GeV and $\tan \beta = 1$, the middle panel is for
$m_A = 500$~GeV and $\tan \beta = 3$, and the right panel is for
$m_A = 500$~GeV and $\tan \beta = 10$, and we assume $m_S=m_T = 1$~TeV in the $h$2MSSM cases.}
\label{XtMSUSY}
\end{center}
\end{figure}

However, the key new point of our analysis is that the required 
values of $M_{SUSY}$ are indeed significantly lower in the $h$2MSSM than in the $h$MSSM. For example,
$M_{SUSY} = 1000$~GeV is now possible for $\tan \beta = 1$ (left panel), 
$M_{SUSY} = 200$~GeV is possible for $\tan \beta = 3$ (middle panel), and even smaller values of $M_{SUSY}$
are possible for $\tan \beta = 10$ (right panel).

Some caveats are in order. As discussed earlier, in this analysis we consider only the
stop contributions to the {$\Delta{\cal M}^2_{22}$ element} in the CP-even Higgs mass matrix. However, as
argued previously, the contributions to other entries in this mass matrix are subdominant,
at least for small $\mu$. Secondly, we have neglected two- and multi-loop effects, but
these should not change our qualitative results. Finally, as also argued previously, the
specifically $N=2$ one-loop corrections due to the adjoint scalar fields are also expected
not to affect significantly our results: for definiteness, we have chosen ${m_S=} m_T = 1$~TeV
in the $h$2MSSM plots in the right panels of Fig.~\ref{XtMSUSY}.

A different way of visualizing our results for the $h$MSSM and $h$2MSSM is shown
in Fig.~\ref{MAtb}. Comparing the two panels, we see that much lower values of $M_{SUSY}$ are
required for the maximal-mixing scenario $X_t = \sqrt{6}M_{SUSY}$ (right panel) than for $X_t = 0$ (left panel).
However, the most striking and novel feature is that, as remarked above, the
$h$2MSSM requires much smaller values of $M_{SUSY}$. When $X_t = 0$ (left panel),
for $\tb \sim 3$ in the $h$MSSM values of $M_{SUSY} \sim 2000$~GeV are required, whereas
$M_{SUSY} > 1000$~GeV are sufficient in the $h$2MSSM. In the maximal-mixing scenario
these values are reduced to $M_{SUSY} \sim 900$~GeV in the $h$MSSM and 
$M_{SUSY} \sim 250$~GeV in the $h$2MSSM.

\begin{figure}[!]
\begin{center}
\includegraphics[width=0.49\textwidth]{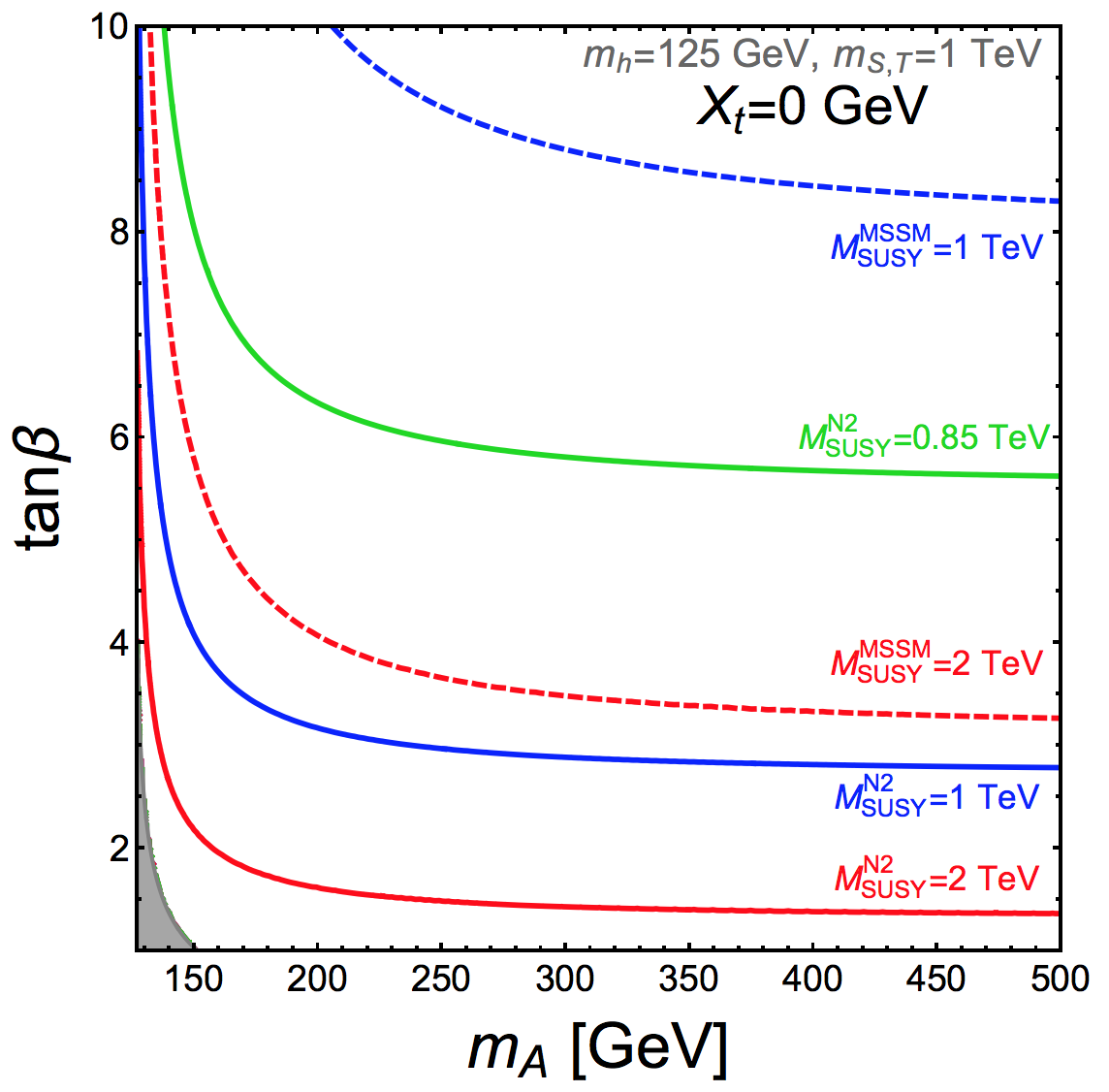} 
\includegraphics[width=0.49\textwidth]{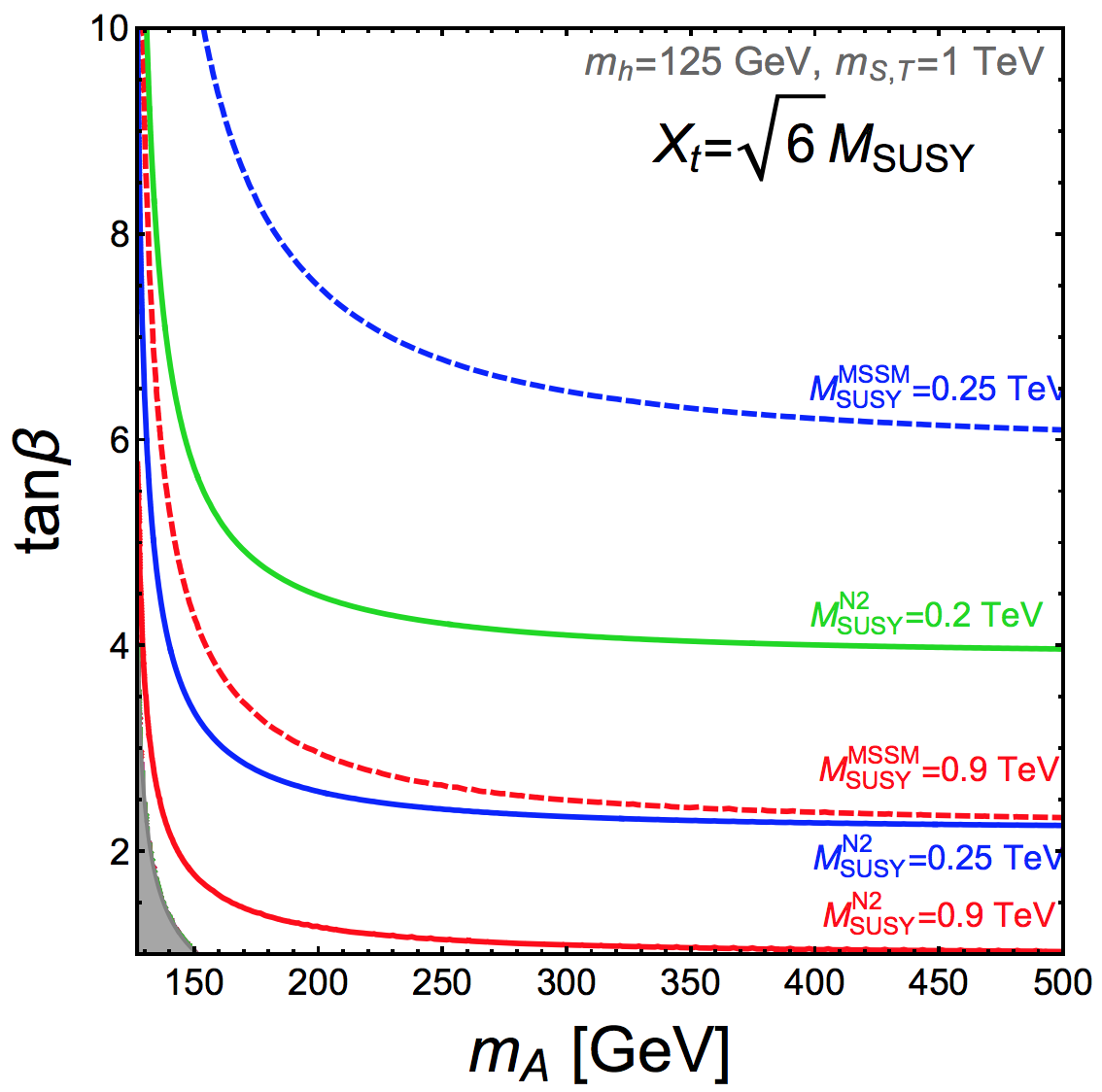}
\vspace{0.5cm}
\caption{\it Contours of $M_{SUSY}$ as functions of $m_A$ and $\tb$ that yield $m_h = 125$~GeV
in the $h$MSSM scenario (dotted lines) and the $N=2$ $h$2MSSM scenario (full lines),
assuming $m_S=m_T = 1$~TeV in the $h$2MSSM cases.
The left panel is for $X_t = 0$, and the right panel is for
the maximal-mixing scenario with $X_t = \sqrt{6} M_{SUSY}$. The grey areas correspond
to the region disallowed in our scenarios,  cf, (\ref{minMA}).}
\label{MAtb}
\end{center}
\end{figure}
%

\section{Constraints from LHC Measurements}\label{sec:couplings}

In light of these differences between the masses and couplings of the Higgs bosons
in the $h$2MSSM and $h$MSSM, we now examine the impacts of LHC constraints in the $(m_A, \tan \beta)$ plane.

\subsection{Constraints from $H/A/H^\pm$ searches}\label{subsec:HAsearches}

Since the mixing angle of the tree-level scalar mass matrix is exactly $\alpha=\beta-\pi/2$ in the $h$2MSSM, 
the heavy Higgs bosons decouple from pairs of gauge bosons at this level, and the loop-induced
$HW^+ W^-$, $HZ^0 Z^0$ and $A Z h$ couplings are relatively small. The limits in the $(m_A, \tan \beta)$
plane of the $N=1$ $h$MSSM coming from $H$ decays to $W^+ W^-$ and $Z^0 Z^0$ and $A$ decay to $Z h$ \cite{hMSSM,Aad:2015pla} are therefore not applicable to the $h$2MSSM. Only the constraints
from $H, A$ and $H^\pm$ couplings to Standard Model fermions are applicable to the $h$2MSSM.
As we have seen, the $H - A$ mass difference is smaller in the $h$2MSSM than in the $h$MSSM,
so the LHC constraints on $A/H \to \tau^+ \tau^-$ are applicable without modification.
This is shown in Fig.~\ref{fit} as a grey excluded region excluding a range of $m_A$ for $\tan \beta \gtrsim 7$.
We do not display the constraint from $H^\pm \to \tau^\pm \nu$ searches, which exclude a small region
at small $m_A$ and large $\tan \beta$ that is contained within the grey area \cite{hMSSM}.

\begin{figure}[t!]
\begin{center}
\includegraphics[width=0.6\textwidth]{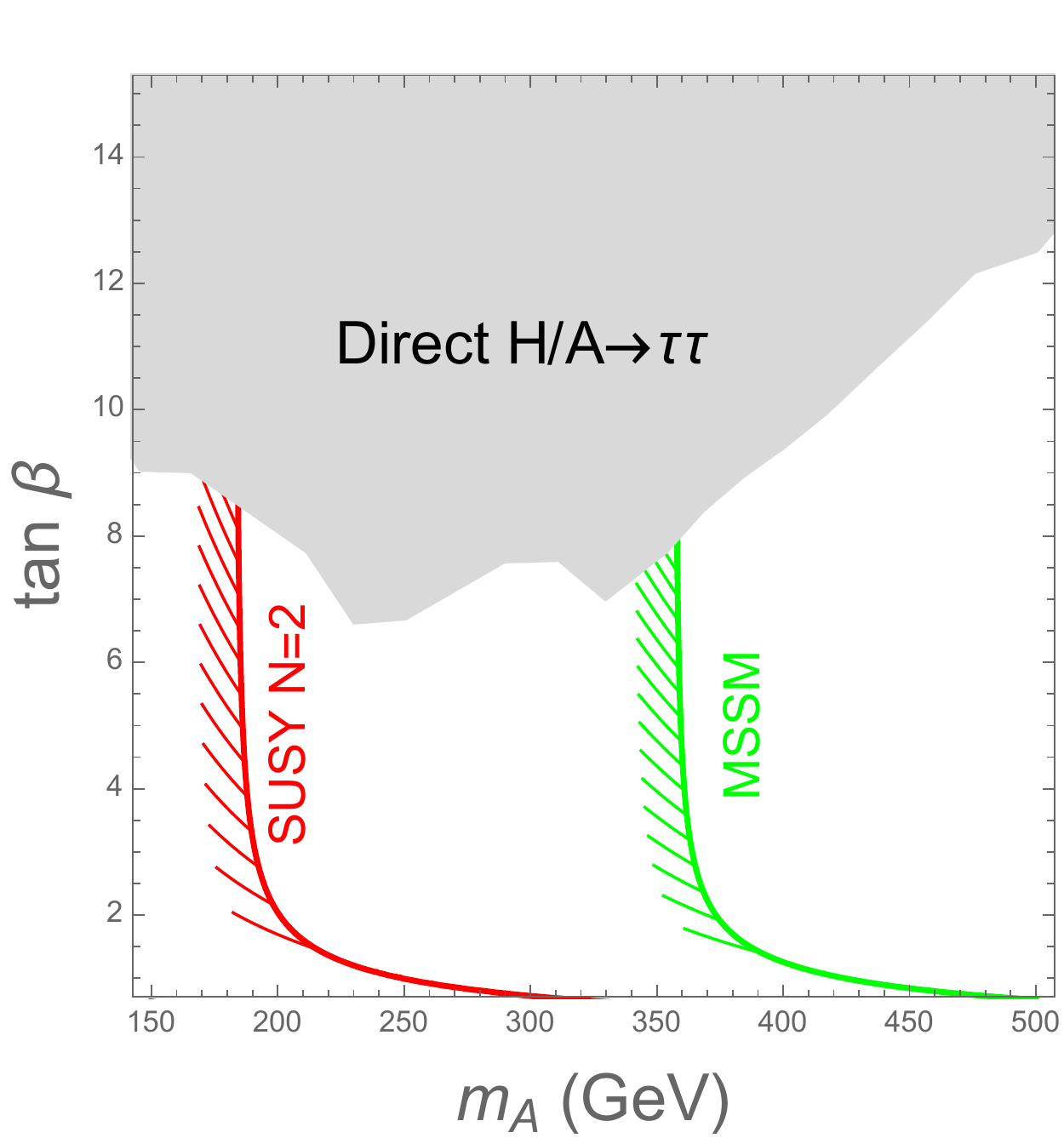} 
\caption{We show in grey the direct exclusion from searches for heavy scalars in the $H/A \to \tau \tau$ final state,
which apply to both the $h$MSSM and the $h$2MSSM. We also show the indirect
bounds from measurements of Higgs couplings to fermions and massive bosons at Run~1 of the LHC for the $h$MSSM (green) 
and $N=2$ $h$2MSSM (red), where the regions to the left of the lines are excluded in each case.}
\label{fit}
\end{center}
\end{figure}
%

\subsection{Constraints from $h(125)$ Coupling Measurements}\label{subsec:run1}

The couplings of the Standard Model-like Higgs boson $h(125)$~\cite{h125} can be analysed using the following
effective field theory (EFT):
\begin{eqnarray} 
{\cal L}_{\text{h-EFT}}  & = &  \ \kappa_V \ g_{hWW} \ h \ W_{\mu}^+ W^{- \mu} + \ \kappa_V \ g_{hZZ} \ h \ Z_{\mu}^0 Z^{0 \mu}
\label{Eq:LagEff}\\ &- &   
  \kappa_t \; y_t\;  h  \bar t_L  t_R  -  \kappa_t \; y_c \; h  \bar c_L  c_R  - 
  \kappa_b \; y_b\;   h  \bar b_L b_R  - \kappa_b \; y_\tau \; h  \bar \tau_L \tau_R 
\  + \ {\rm h.c.} \, , \nonumber \label{Lagfit}
\end{eqnarray}
where $y_{t,c,b,\tau}=m_{t,c,b,\tau}/v$ are the Standard Model Yukawa couplings in
the mass eigenbasis, the subscripts $L/R$ label the left and right chirality states of the fermions, and
we consider only the fermions with the largest couplings to the Higgs boson.  The quantities
$g_{hWW} = 2m^2_W/v$ and $g_{hZZ} = m^2_Z/v$ are the couplings of $h$ to the electroweak gauge bosons,
and $v$ is the vacuum expectation value of the Higgs field.  The parameters $\kappa_{X}$ are the free parameters of
this EFT.  

These parameters can be constrained using the Higgs signal strengths in
various channels, denoted by $XX$:
\beq
\mu_X \equiv \frac{\sigma( pp \to h) \times {\rm BR}(h \to XX)}{\sigma(
pp \to h)_{\rm SM}  \times {\rm BR}(h \to XX)_{\rm SM}} \, ,
\eeq
as measured in all the Higgs production/decay channels available from the LHC Run 1.
A full analysis requires performing an appropriate three-parameter fit in the three-dimensional 
$(\kappa_V,\kappa_t, \kappa_b)$ space, where we assume that $\kappa_c\!=\!\kappa_t$, $\kappa_\tau\!=\!\kappa_b$,
which is consistent with the current experimental accuracies, and
$\kappa_V\!=\!\kappa_W\!=\!\kappa_Z$,  the custodial symmetry relations that should hold to
 a good approximation in the supersymmetric models of interest.

In our two supersymmetric models, the $N=1$ MSSM and the $N=2$ $h$2MSSM scenario,
the $\kappa$ parameters take the following similar forms:
\begin{eqnarray}  
\kappa_V   =   \sin(\beta- \alpha)  \ , \ \  \kappa_t   =  
\frac{\cos  \alpha}{\sin\beta} \ , \ \  \kappa_b   =   -
\frac{\sin  \alpha}{\cos\beta}  \label{kappaSUSY} 
\end{eqnarray}
where $\alpha$ is the rotation angle that diagonalizes the Higgs mass-squared matrix in the $h$MSSM or $h$2MSSM,
respectively, after including the dominant one-loop radiative corrections as discussed above.
The expressions (\ref{kappaSUSY}) do not include the effects of subdominant loop corrections,
which may not be negligible if the supersymmetric particles are not very heavy, in which case there are direct radiative
corrections to the Higgs couplings that are not contained in the expression of the mass matrix. 
We neglect such possible effects in the present study.

At tree level, $\alpha$ only depends on two unknown quantities, namely $\tan\beta$ and $m_A$.
Moreover, only two of the three quantities $\kappa_V$, $\kappa_t$ and $\kappa_b$ are independent.
This is still the case when we include the dominant one-loop radiative corrections and fix $m_h=125$~GeV
as discussed above. In both the $h$MSSM and the $h$2MSSM we can derive $\kappa_V(\tan\beta,m_A)$,
$\kappa_t(\tan\beta,m_A)$ and $\kappa_b(\tan\beta,m_A)$ for any pair of values of $(\tan\beta,m_A)$. 

The values may be derived by plugging the explicit expressions for $\alpha_{MSSM}$ in 
(\ref {hMSSM-output}) and $\alpha_{N2}$ in (\ref{h2MSSM-output})
into (\ref{kappaSUSY}).
Alternatively, one can proceed directly from the MSSM or $N=2$ mass-squared matrix,
associating the mass eigenvalue $m_h$ with the normalized eigenvector $V_h=(V_{h 1},V_{h 2})$
such that the physical field is $h=V_{h i}H_i$ with $i=1,2$ and
the mass eigenvalue $m_{H}$ with the normalized eigenvector $V_H=(V_{H 1},V_{H 2})$
such that the physical field is $H=V_{Hi}H_i$ with $i=1,2$.
We then have
\bea
\kappa_t  &=&  \frac{1}{\sin\beta}\, V_{h2}(\tan\beta,m_A) \, , \quad \kappa_b  =  \frac{1}{\cos\beta} \, V_{h1}(\tan\beta,m_A) \, , \nonumber \\
\kappa_V  &=&  \sin\beta \, V_{h2}(\tan\beta,m_A) + \cos\beta V_{h1}(\tan\beta,m_A) \, .
\eea
In terms of $\tan\beta$ we find
 \bea
\kappa_t  &=&  \frac{\sqrt{1+\tan^2\beta}}{\tan\beta}\, V_{h2}(\tan\beta,m_A) \, ,\quad
\kappa_b =  \sqrt{1+\tan^2\beta} \, V_{h1}(\tan\beta,m_A) \, , \nonumber \\
\kappa_V &=&  \frac{1}{\sqrt{1+\tan^2\beta}}(\tan\beta \, V_{h2}(\tan\beta,m_A) +  V_{h1}(\tan\beta,m_A)) \, ,
\eea
where in the case of the $h$MSSM:
\bea
V_{h2}^{MSSM}(\tan\beta,m_A) & = & \frac{1}{\sqrt{1+\left(  \frac{(m_A^2+m_Z^2)\tan\beta}{m_Z^2-m_h^2(1+\tan^2\beta)+m_A^2\tan^2\beta}\right)^2}} \, , \\
V_{h1}^{MSSM}(\tan\beta,m_A) & = & \frac{(m_A^2+m_Z^2)\tan\beta}{m_Z^2-m_h^2(1+\tan^2\beta)+m_A^2\tan^2\beta} V_{h2} \, ,
\eea
and in the case of the $N=2$ $h$2MSSM: 
\bea
V_{h2}^{N2}(\tan\beta,m_A) & = & \frac{1}{\sqrt{1+\left(  \frac{(m_A^2-m_Z^2)\sin2\beta}{m_A^2-2m_h^2+m_Z^2+(m_Z^2-m_A^2)\cos2\beta}\right)^2}} \, , \\
V_{h1}^{N2}(\tan\beta,m_A) & = & \frac{(m_A^2-m_Z^2)\sin2\beta}{m_A^2-2m_A^2+m_Z^2+(m_Z^2-m_A^2)\cos2\beta} V_{h2} \, .
\eea
These results can be used to apply the constraints on Higgs couplings derived from a combination of CMS and ATLAS data at
Run1~\cite{Khachatryan:2016vau}. In particular, the analysis relevant to constraining the $h$MSSM
and $h$2MSSM scenarios tests for deviations from the Standard Model in couplings to up- and down-type quarks 
and to vector bosons via the ratios $\lambda_{du}$ and $\lambda_{Vu}$:
\begin{eqnarray}
\lambda_{du} &=& \frac{\kappa_d}{\kappa_u} = 0.92^{+0.12}_{-0.12} \, ,  \nonumber \\
\lambda_{Vu} &=& \frac{\kappa_V}{\kappa_u} = 1.00^{+0.13}_{-0.12} \, . 
\end{eqnarray}
The results of this fit are shown in Fig.~\ref{fit}, where the excluded region in the $h$MSSM
lies to the left of the green line, whereas in the $N=2$ case the bounds (in red) are very much weakened.

We conclude from Fig.~\ref{fit} that $m_A \gtrsim 200$~GeV is allowed in the $h$2MSSM for $\tb \in (2, 8)$,
whereas $m_A \gtrsim 350$~GeV would be required in the $h$MSSM.

\subsection{Constraints from $\Gamma(h \rightarrow gg, \gamma\gamma)$}\label{subsec:hgamgam}

We now analyze the corrections to the couplings of the SM-like Higgs boson to gluons and photons that arise at the loop level,
and the corresponding constraints on the $h$MSSM and $h$2MSSM.

The decay width of the Standard Model-like $h(125)$ into pairs of gluons and photons can be expressed 
as \cite{Gunion:1989we,Djouadi:1996pb}:
\beq
\Gamma (h \rightarrow g g) = \frac{G_F \alpha_s^2 m_{h}^3}{64 
\sqrt{2} \pi^3 } \left| \, \sum_i A^{gg}_i (\tau_i) \, \right|^2 \, , \; 
\Gamma (h \rightarrow \gamma \gamma) = \frac{G_F \alpha^2 m_{h}^3}{128 
\sqrt{2} \pi^3 } \left| \, \sum_i A^{\gamma\gamma}_i (\tau_i) \, \right|^2 \, ,
\eeq
where the variable $\tau_i \equiv m_{h}^2/4m_i^2$, $m_i$ being the mass of the particle propagating in the loop.
In the case of the loops for the $hgg$ coupling, whereas one has only contributions from quarks in the Standard
Model, in the MSSM additional contributions are provided by the scalar partners of those quarks. 
The normalized amplitudes of these two contributions are
\bea
A^{gg}_f = g_{hff} \, F_{1/2} (\tau_f) \, \quad , A^{gg}_{\tilde{f_i}} = g_{h \tilde{f}_i \tilde{f}_i } 
\frac{ M_Z^2} {m_{\tilde{f}_i}^2 } \, F_{0} (\tau_{\tilde{f}_i}) \, .
\label{Agg}
\eea
In the case of the loop for the $h \gamma \gamma$ coupling, in the Standard Model the $W$ boson and 
charged fermions are the only contributors, whereas in the MSSM there are additional contributions
from the two chargino fermionic fields, the scalar partners of the fermions and the charged Higgs boson.
The normalized amplitudes of these contributions are
\beq
A^{\gamma\gamma}_W &=& g_{\Phi WW} \, F_1 (\tau_W) \, , \quad
A^{\gamma\gamma}_f = N_c Q_f^2 g_{\Phi ff} \, F_{1/2} (\tau_f) \, , \quad A^{\gamma\gamma}_{\chi_i} = g_{\Phi \chi_i^+ \chi_i^-} \frac{M_W}{m_{\chi_i}} \, F_{1/2} 
(\tau_{\chi_i})  \, , \nonumber \\
A^{\gamma\gamma}_{\tilde{f_i}} &=& N_c Q_f^2 g_{\Phi \tilde{f}_i \tilde{f}_i } 
\frac{ M_Z^2} {m_{\tilde{f}_i}^2 } \, F_{0} (\tau_{\tilde{f}_i}) \, , \quad A^{\gamma\gamma}_{H^\pm} = g_{\Phi H^+ H^-}\frac{M_W^2}{M_{H^\pm}^2} \, F_0 (\tau_{H^\pm}) \, , 
\label{Agamgam}
\eeq
where $N_c$ is the color factor and $Q_f$ the electric charge of the fermion or sfermion 
in units of the proton charge. 

The spin 1, 1/2 and 0 amplitudes are \cite{Gunion:1989we}
\beq
F_1(\tau) &=& [2\tau^2+3 \tau +3(2\tau-1) f(\tau)]/\tau^2 \, , \nonumber \\
F_{1/2}(\tau) &=& - 2 [\tau+(\tau-1)f(\tau)]/\tau^2 \, , \nonumber \\
F_0 (\tau) &=& [\tau -f(\tau)]/\tau^2 \, , 
\eeq
with the function $f(\tau)$ defined as
\begin{equation}
f(\tau) = \left\{ \begin{array}{ll} 
{\rm arcsin}^2 \sqrt{\tau} & \tau \leq 1 \, ,  \\
-\frac{1}{4} \left[ \log \frac{1 + \sqrt{1-\tau^{-1} } }
{1 - \sqrt{1-\tau^{-1}} } - i \pi \right]^2 \ \ \ & \tau >1 \, .
\end{array} \right. 
\end{equation} 
The amplitudes are real when $m_h < 2m_i$, but are complex above that threshold. 
In the regime $\tau \ll 1$, i.e., heavy masses in the loop, the amplitudes reach asymptotic 
values
\beq
F_{1} \rightarrow +7 \ \ , \ \ F_{1/2} \rightarrow -\frac{4}{3} \ \ {\rm and}
\ \ F_{0} \rightarrow - \frac{1}{3} .
\eeq
Standard Model particle loops give finite contributions in the heavy-mass limit, 
whereas the new supersymmetric contributions decouple in the limit of large mass, since their amplitudes $A_i$ are divided by their masses. 

As we have discussed in the previous Section, the top quark superpartners are responsible for a substantial shift
in the tree-level Higgs mass of $\sim 34$ GeV in the $h$2MSSM (and more in the $h$MSSM). 
We will focus in the following on the loop-level correction to the $hgg$ and $h\gamma\gamma$ couplings due to the stops, neglecting other potential supersymmetric contributions.

The loop-level corrections from stops to Higgs production via gluon-gluon fusion
and to $h \to \gamma \gamma$ decay are given, respectively, by
\beq
\frac{\sigma(gg \rightarrow h)}{\sigma^{SM}(gg \rightarrow gg)} \simeq \frac{\Gamma(h \rightarrow gg)}{\Gamma^{SM}(h \rightarrow gg)} \simeq |\kappa_g|^{2} , \qquad 
\frac{\Gamma(h \rightarrow \gamma\gamma)}{\Gamma^{SM}(h \rightarrow \gamma\gamma)} \simeq |\kappa_\gamma|^{2} \, ,
\eeq
with
\beq
\kappa_g = 1 + \frac{A^{gg}_{\tilde{t}_1}+A^{gg}_{\tilde{t}_2}}{\sum_{i \in SM} A^{gg}_i} , \qquad 
\kappa_\gamma = 1 + \frac{A^{\gamma\gamma}_{\tilde{t}_1}+A^{\gamma\gamma}_{\tilde{t}_2}}{\sum_{i \in SM} A^{\gamma\gamma}_i} \, .
\eeq
It has been shown that, to a good approximation \cite{Espinosa:2012in}, $\kappa_{g,\gamma}$ reduce to
\beq
\kappa_g \simeq 1 + \frac{A_{\tilde{t}}}{\sum_{i \in SM} A^{gg}_i} , \qquad 
\kappa_\gamma \simeq 1 + \frac{N_c Q_{\tilde{t}}^2 A_{\tilde{t}}}{\sum_{i \in SM} A^{\gamma\gamma}_i} \, ,
\eeq
where
\beq
A_{\tilde{t}}=-\frac{1}{3}\left( \frac{m_t^2}{m_{\tilde{t}_1}^{2}}+\frac{m_t^2}{m_{\tilde{t}_2}^{2}}-\frac{1}{4}\sin^2(2\theta_t)\frac{(m_{\tilde{t}_2}^{2}-m_{\tilde{t}_1}^{2})^2}{m_{\tilde{t}_1}^{2}m_{\tilde{t}_2}^{2}} \right) \, ,
\eeq
with $\theta_t$ the mixing angle of the scalar mass matrix. We remind the reader that the physical stop masses are
\beq
m^2_{\tilde{t}_1 , \tilde{t}_2} = m_t^2+ \frac{1}{2} \left[ 
m_{\tilde{t}_L}^2 + m_{\tilde{t}_R}^2 \mp \sqrt{( m_{\tilde{t}_L}^2 
-m_{\tilde{t}_R}^2)^2+ (2 m_t X_t)^2 } \, \right] \, ,
\eeq
where $X_t = A_t - \mu/\tan\beta$, and $A_t$, $m_{\tilde{t}_R}$ and $m_{\tilde{t}_L}$ are parameters of the soft
supersymmetry-breaking Lagrangian, and the squark mixing angle, $\theta_t$, is defined by
\beq
\sin 2\theta_t = \frac{2 m_t X_t}{m_{\tilde{t}_1}^2 - 
m_{\tilde{t}_2}^2} \ \ , \ \ \cos 2 \theta_t = \frac{m_{\tilde{t}_L}^2 - 
m_{\tilde{t}_R}^2 } {m_{\tilde{t}_1}^2 - m_{\tilde{t}_2}^2} \, .
\eeq
The stop sector can be parametrised by the three inputs $m_{\tilde{t}_L}$, $m_{\tilde{t}_R}$ and $X_t$ or, alternatively,
by the physical stop masses $m_{\tilde{t}_1}$, $m_{\tilde{t}_2}$ and $X_t$.
If the mixing parameter is large, the two stop masses are strongly split, $m_{\tilde{t}_1} \ll m_{\tilde{t}_2}$,
and the ${\tilde{t}_1}$ has a large coupling to the $h(125)$ state, $g_{\tilde{t}_1\tilde{t}_1} \propto m_t X_t$.

If we consider the  $[m_{\tilde t_1}, m_{\tilde t_2}]$ plane for fixed values of $m_A$ and $\tan\beta$,
we can fix $X_{t}^{2}$ by the requirement that $m_h=125$ GeV when just the dominant stop
contributions to the radiative corrections in the MSSM Higgs sector are considered~\cite{Djouadi:2015aba}. In this case,
the shift of the Higgs mass is given by (\ref{stoptoploop}) and (\ref{STbit}) in the $h$MSSM and $h$2MSSM, respectively. 
There are at most two solutions for $X_t^{2}$, denoted by $|X_t^{max}|$ and $|X_t^{min}|$. Having traded
the stop mixing parameter by the requirement $m_h=125$~GeV, we can now compute the couplings between the stops and the 
$h(125)$ and then $\kappa_{g,\gamma}$.

The available experimental constraints on $\kappa_{\gamma}$ are shown  in green (red) for the $h$MSSM ($h$2MSSM) in 
Fig.~\ref{mstopskappagam} for $m_A=500$~GeV and $\tan\beta=1.5$ (upper panels), $\tan\beta=5$ (middle panels)
and $\tan\beta=10$ (lower panels). In the case of the h$2$MSSM, we always consider a generic common adjoint scalar mass
$m_S = m_T = 1$~TeV.
The constraints on $\kappa_{g}$ are less severe than those on
$\kappa_\gamma$, so we do not display them in Fig.~\ref{mstopskappagam}.

The Higgs mass requirement has, in general, zero, one or two solutions for $X_t^{2}$,
and it is possible that one or more of them might be in conflict with the constraint coming from the soft masses:
\beq
(m_{\tilde{t}_L}^2 -m_{\tilde{t}_R}^2)^2 = (m_{\tilde{t}_1}^2 -m_{\tilde{t}_2}^2)^2 - (2 m_t X_t)^2 \, ,
\eeq
from which we can derive the maximum allowed value for $X_t$, $|X_{t}^{soft}|$, which is given by
\beq
X_{t}^{soft, 2}=\frac{(m_{\tilde{t}_1}^2 -m_{\tilde{t}_2}^2)^2}{4m_{t}^{2}} \, .
\eeq
When scanning the $(m_{\tilde{t}_{1}}$, $m_{\tilde{t}_2})$ plane, we must ensure that our solutions in $X_t$ are below this maximal value.
The grey regions in Fig.\ref{mstopskappagam} with dotted (full) border contours are forbidden by this
consideration in the case of the $h$MSSM ($h$2MSSM). There are no values of $X_t$
able to accommodate $m_h=125$~GeV in the hMSSM (h$2$MSSM) in the regions at low
$m_{\tilde{t}_{1}}$ and/or $m_{\tilde{t}_2}$ that are shaded yellow (blue). 

The left panels of Fig.~\ref{mstopskappagam} consider the maximal value of $X_t$ allowing $m_h=125$~GeV,
including the case where there is only one possible choice for $X_t$.
The right panels of Fig.~\ref{mstopskappagam} consider the minimal value of $X_t$ allowing $m_h=125$~GeV,
including the case where there is only one possible choice for $X_t$.
This explains the particular shape of the grey region for relatively high stop masses.

The current constraints on $\kappa_{g,\gamma}$ in the $h$MSSM and the $h$2MSSM
are outlined in green (red) in Fig.~\ref{mstopskappagam}. We see that they are generally weak. Indeed, 
for $m_A=500$ GeV and $\tan\beta=1.5$ (top two panels) there is no constraint at all. 
However, for higher values of $\tan\beta$ (middle and bottom panels) these constraints do exclude some scenarios with low
supersymmetry-breaking scales.

\begin{figure}[!]
\vspace{-1cm}
\begin{center}
\includegraphics[width=0.42\textwidth]{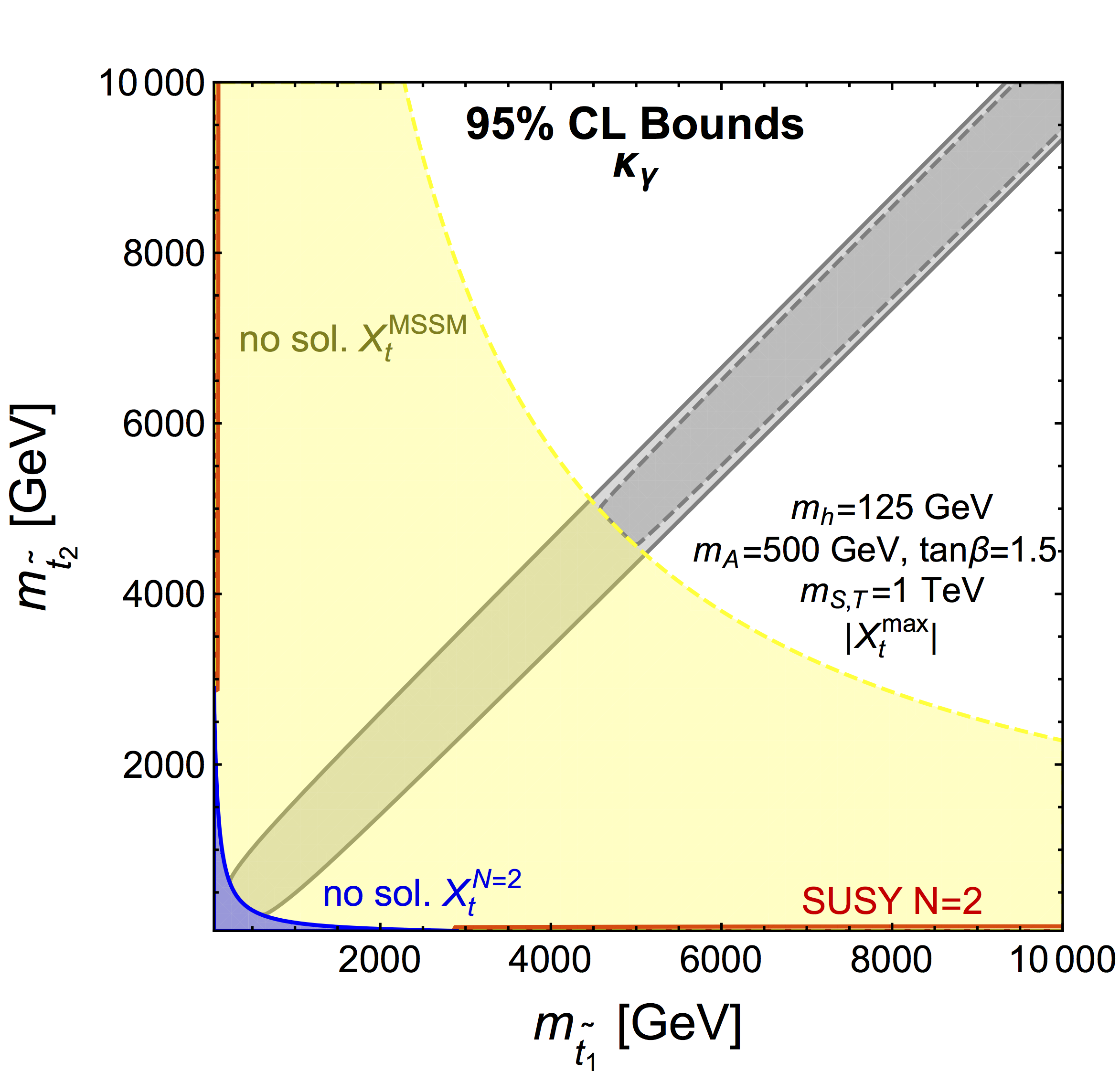} 
\includegraphics[width=0.42\textwidth]{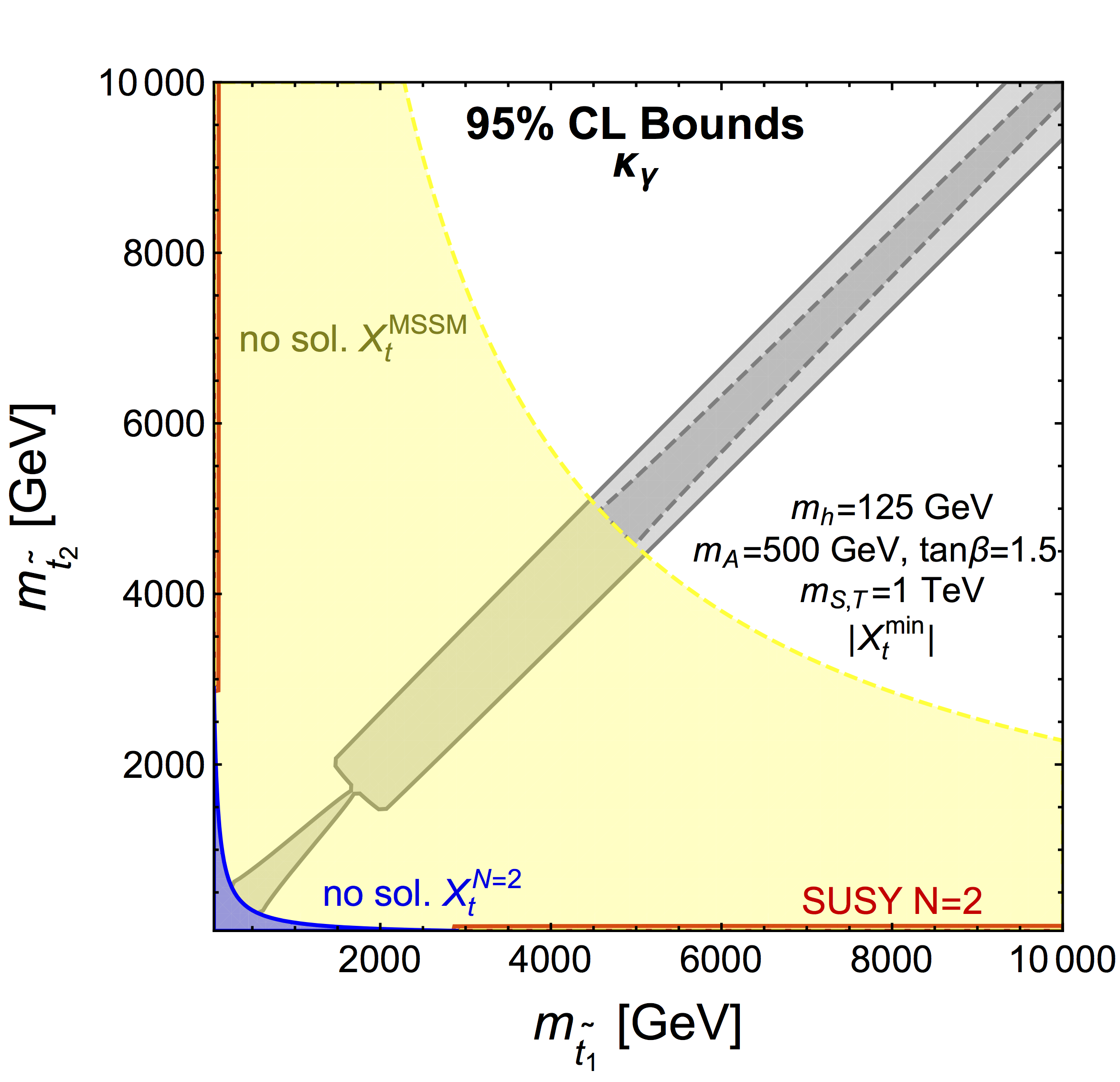} \\ \vspace{-0.5cm}
\includegraphics[width=0.42\textwidth]{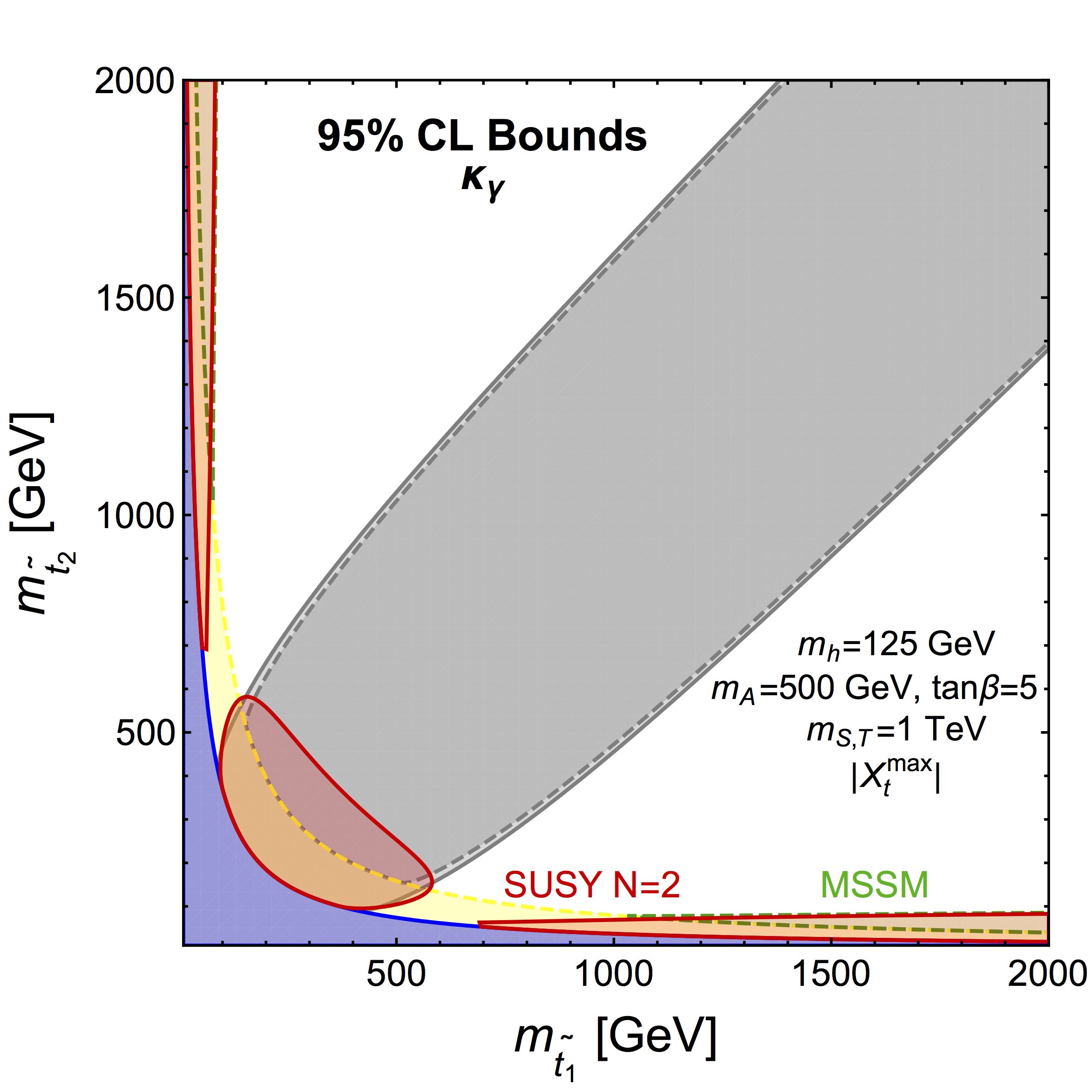} 
\includegraphics[width=0.42\textwidth]{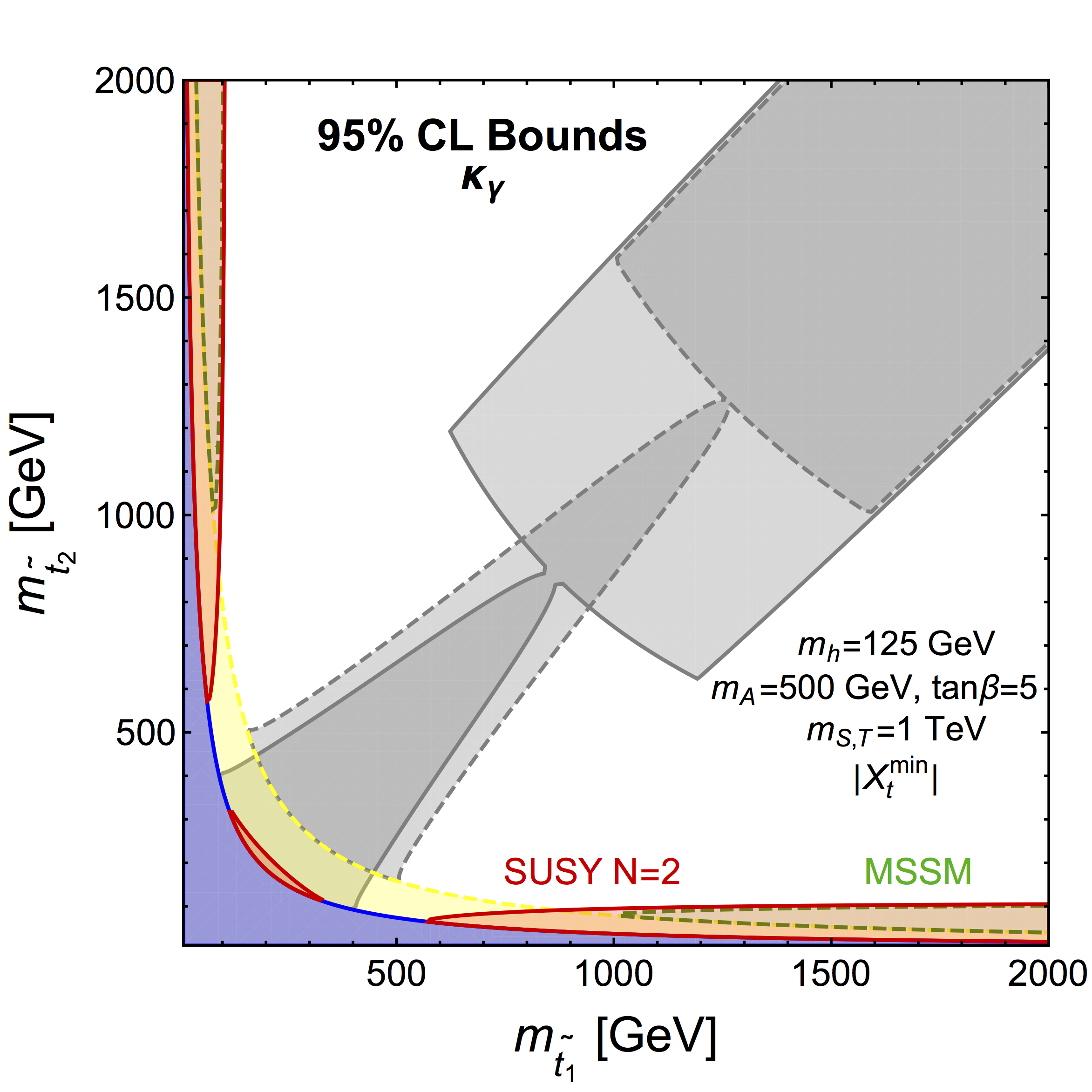} \\ \vspace{-0.5cm}
\includegraphics[width=0.42\textwidth]{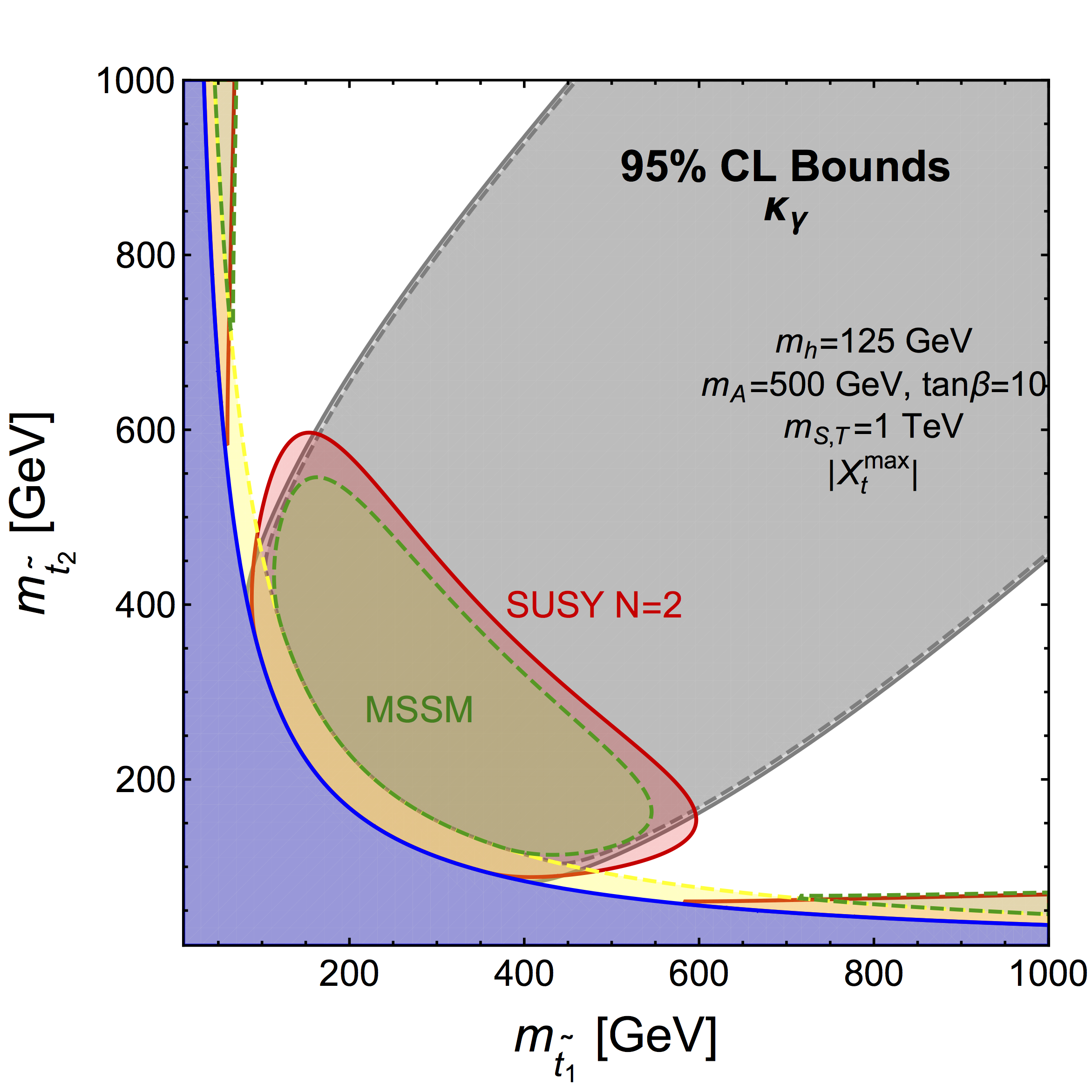} 
\includegraphics[width=0.42\textwidth]{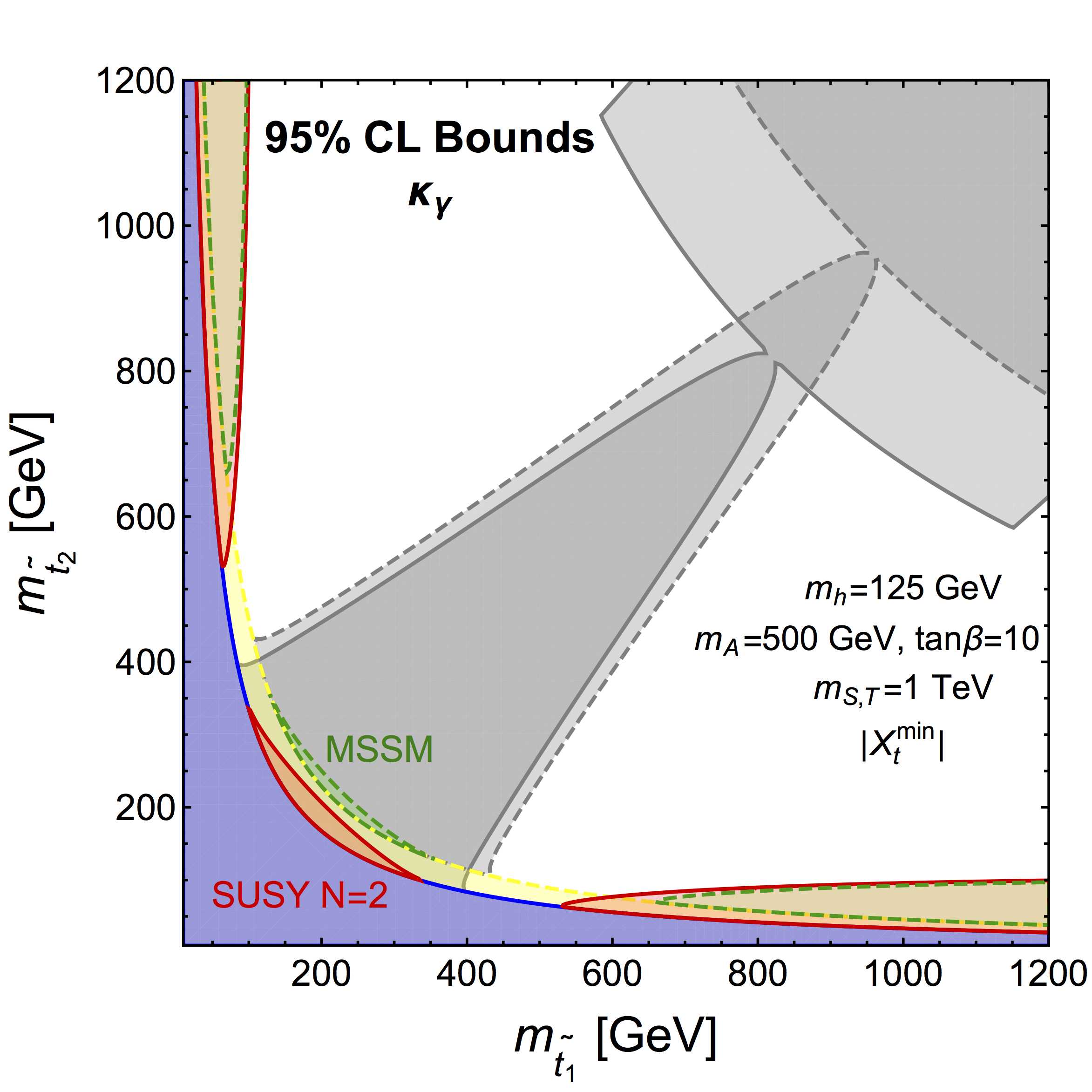} 
\vspace{-0.3cm}
\caption{\it 
Compilation of the constraints in $(m_{\tilde{t}_{1}}$, $m_{\tilde{t}_2})$ planes fixing $X_t$ so as to
obtain $m_h = 125$~GeV in the $h$MSSM and $h$2MSSM,
assuming $M_A=500$ GeV and $\tan\beta=1.5$ (top panels), $\tan\beta=5$ ( middle panels) and 
$\tan\beta=10$ (bottom panels). In the case of the h$2$MSSM we assume $m_S=m_T=1$ TeV. 
For any given pair of stop masses, the $m_{h}=125$ GeV requirement allows
at most two solutions for the stop mass mixing parameter,$X^2_t$. The left (right) panels
correspond to the maximal (minimal) solution, $|X_t^{max}|$ ($|X_t^{min}|$).
The grey regions bounded by dashed (full) contours are disallowed by the mixing hypothesis in the $h$MSSM ($h$2MSSM).
Regions where there are no values of $X_t$ that yield $m_h=125$~GeV in the hMSSM (h$2$MSSM) are shaded yellow (blue).
The regions inside the red (green) contours are forbidden by the LHC $h \to \gamma \gamma$ constraint
in the $h$2MSSM ($h$MSSM).}
\label{mstopskappagam}
\end{center}
\end{figure}
%

\subsection{Anomalous $h(125)$ Couplings}\label{subsec:anom}

In addition to these modifications of the $h$ couplings measured in Higgs production and decay,
integrating out the heavy scalars can also induce anomalous couplings of the Higgs to vector bosons with non-standard momentum dependence. 
One can parametrize these effects in the coupling of the Higgs to two $W$ bosons as follows~\cite{Artoisenet:2013puc}:
\bea
\label{LWWh}
\Delta \mathcal{L}_{W} &=& -\frac{g^{(1)}_{hWW}}{2}\, W^{\mu\nu}W^{\dagger}_{\mu\nu} h - \left[ g^{(2)}_{hWW}\,W^{\nu}\partial^{\mu}W^{\dagger}_{\mu\nu}h + \mathrm{h.c.}\right] + g^{(3)}_{hWW} \,W^{\mu}W^{\dagger}_{\mu}h \, .
\eea
We note that the coupling $g^{(3)}$ causes a shift in the usual Standard Model coupling structure. Indeed,
the interpretation of the Higgs data described by the Lagrangian (\ref{Lagfit})  corresponds to $g^{(3)}= (\kappa_V-1) g_{hWW}$
and setting $g^{(1,2)}$ to zero. However, with more precise measurements of differential distributions in Run~2 one may be able to 
disentangle different Lorentz structures, which could give a handle for discriminating between an anomaly due to the MSSM and an 
underlying $N=2$ supersymmetric structure.

\begin{figure}[t!]
\begin{center}
\includegraphics[width=0.4\textwidth]{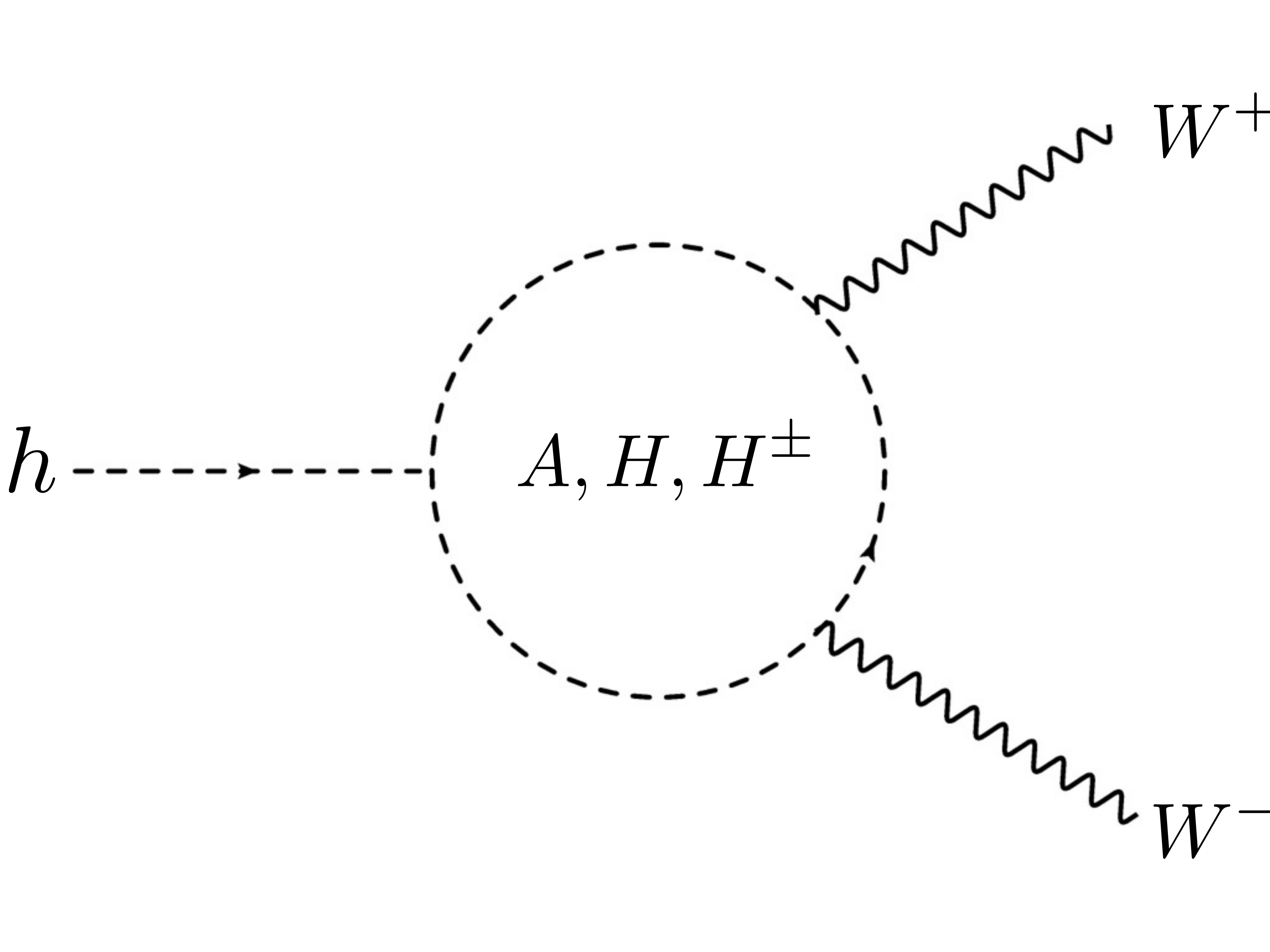} 
\includegraphics[width=0.4\textwidth]{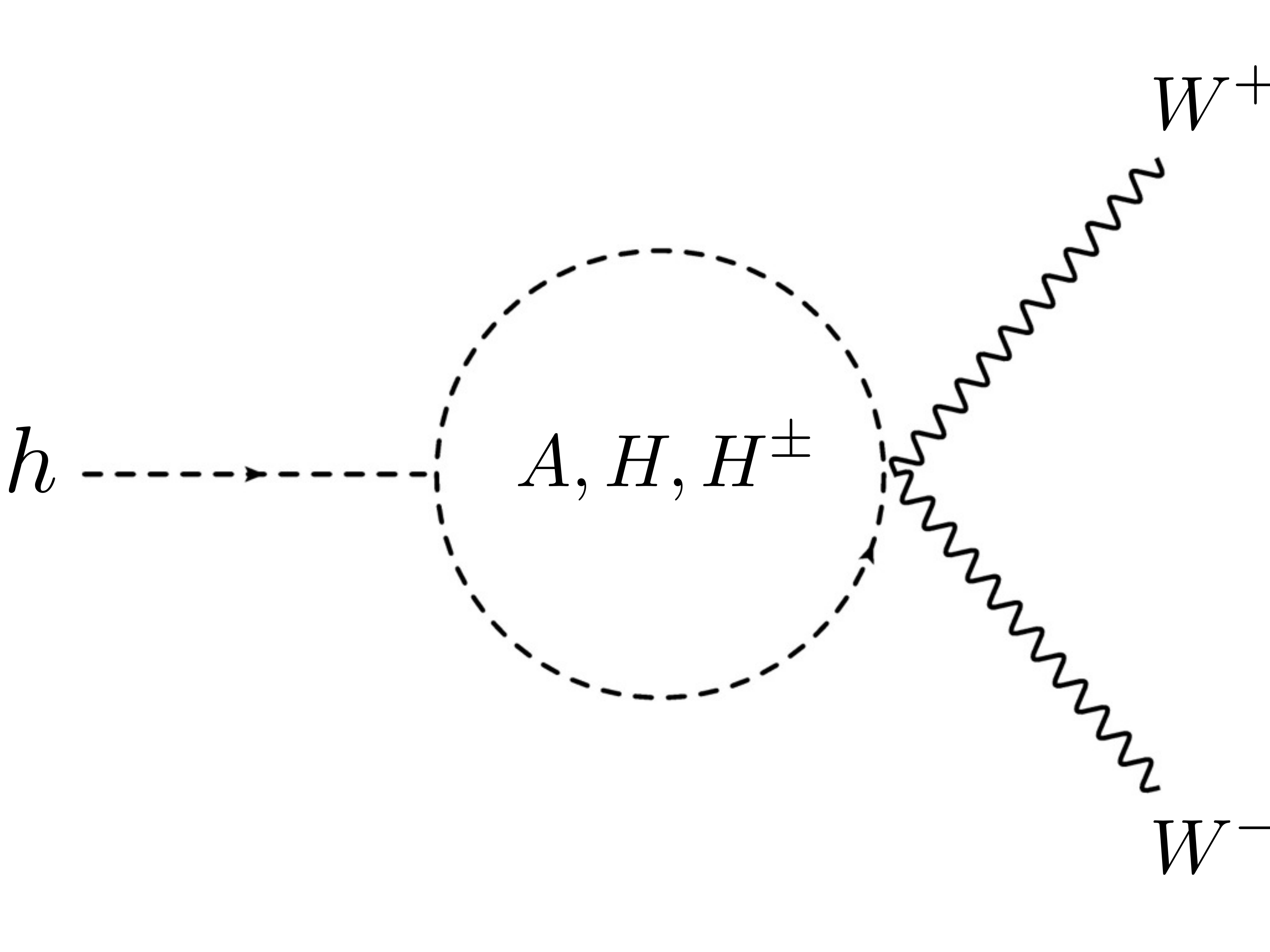} 
\caption{\it Loop contributions of the heavy scalars to anomalous $h(125)$ couplings.}
\label{figloop}
\end{center}
\end{figure}

Generic expressions for the effects of one-loop scalar contributions to Higgs anomalous couplings can be
found in~\cite{Gorbahn:2015gxa}. These correspond to integrating out the heavy MSSM Higgs bosons
$A$, $H$ and $H^{\pm}$ in loops, as shown in Fig.~\ref{figloop}. It is important to note that electroweak precision tests, 
particularly the constraints from the  $S$ and $T$ parameters, require the values of $m_A$, $m_H$ and $m_{H^\pm}$ 
to be relatively close to each other. In particular, in a 2HDM the expression for $\Delta S$ and $\Delta T$ is given by~\cite{Gorbahn:2015gxa} 
\bea
\Delta S &=& - \frac{g_2^2\, s_W^2 (1- x_A + 1- x_0)}{96\, \pi^2 \,\alpha_{\mathrm{EM}}}  \, , \nonumber \\
\Delta T &=& \frac{m_{H^\pm}^2 (1-x_A) (1-x_0) }{48 \, \pi^2 \,v^2 \,\alpha_{\mathrm{EM}}} \, ,
\eea
where we define the splittings among the heavy scalars by the quantities $x_{0,A}$:
\bea
x_0 \equiv \frac{m^2_{H}}{m^2_{H^{\pm}}} \ , \, x_A \equiv \frac{m^2_{A}}{m^2_{H^{\pm}}} \ .
\eea 
and have expanded at linear order in  $1-x_{0,A}$. As the splittings in this model are small, imposing the current best fit values from the global analysis of the GFitter  group~\cite{Gfitter} does not restrict further the parameter space of ($m_A$, $\tan \beta$) from the Higgs coupling constraints. Indeed, $\Delta S$, $\Delta T\sim 10^{-2}$ for $m_A \gtrsim$ 100 GeV.

In this approximation, one can find compact expressions for the
anomalous Higgs couplings:
\bea
\label{gWWh1}
g^{(1)}_{hWW} & = &\frac{-g_2^2\,v}{192 \, \pi^2\,m^2_{H^{\pm}}} \left[\frac{g_0 + g_A + 2 g_+}{2}+ (1-x_0)\frac{4g_0 + g_+}{10} + (1-x_A)\frac{4g_A + g_+}{10} 
\right] \, , \nonumber  \\
g^{(2)}_{hWW} & = & \frac{g_2^2\,v}{192 \, \pi^2\,m^2_{H^{\pm}}} \left[(1-x_0)\frac{g_0 - g_+}{20} + (1-x_A)\frac{g_A - g_+}{20}
\right] \, , \nonumber  \\
\label{gWWh3}
g^{(3)}_{hWW} & = & \frac{g_2^2\,v}{192 \, \pi^2} \left[(1-x_0) (g_+- g_0) + (1-x_A) (g_+- g_A)
\right] \, .
\eea
Here $g_{0,A,+}$ denote the trilinear scalar couplings, $g_0 \equiv g_{hHH}/v$, $g_A \equiv g_{hAA}/v$ and $g_+ \equiv g_{hH^+H^-}/v$. These expressions are generic in a 2HDM model as long as the expansion in $x_{0,A}$ is justified.  

The values of the splittings in the MSSM and its $N=2$ extension can be obtained by inspecting
(\ref{massmatrixMSSM}) and (\ref{massmatrixN2}), respectively.  In the $N=2$ case, one finds $x_0=x_A\simeq 1- m_W^2/m_A^2$. 

Turning now to the trilinear Higgs couplings, we note that the new $N=2$ term in the scalar potential in (\ref{pot})
does not contribute, so the analytical formulae for the trilinear couplings are the same as in the $N=1$ MSSM, 
see, e.g.,~\cite{Djouadi:2005gj}. 
Therefore, at leading order in $m_W^2/m_A^2$, the effect of integrating out the heavy scalars in the $N=2$
extension of the MSSM is to generate anomalous couplings of the Higgs to vector bosons of the type $g^{(1)}_{hWW}$,
namely a Higgs coupling to the square of the gauge field strength with magnitude
\bea
g^{(1)}_{hWW} & = &\frac{-g_2^2\,v}{192 \, \pi^2\,m^2_{A}} \left[ 1+ 2 c_W^2 - 3 \, \frac{m_h^2-\epsilon}{m_Z^2}  \right]  \ .
\eea
Bounds on effective operators in an Effective Field Theory approach from Higgs data using differential 
distributions~\cite{Ellis:2014jta,Ellis:2014dva} can be used in our case by noting that the anomalous couplings 
are related to operators defined there by~\cite{Gorbahn:2015gxa}
\bea
g_{hWW}^{(1)} &=& \frac{2 g_2}{m_W} \bar c_{HW} \, , \\
 g_{hWW}^{(2)} &=& \frac{g_2}{2\, m_W} \Big[ \bar c_{W} + \bar c_{HW} \Big] \, .
\eea 
This leads to a specific relation among the operators, namely $\bar c_W = - \bar c_{HW}$ for this model. 

A global fit to Higgs and electroweak boson properties in this particular case was made in~\cite{Ellis:2014jta}, 
leading to a bound from the Run~1 data: $\bar c_{HW} \in (0.0004,0.02)$, which places no useful constraint on $m_A$
currently, as compared with the bounds on total rates discussed before. However, this situation may change with the
advent of Run~2 and subsequent Higgs data.
 
\section{Conclusions}\label{sec:concl}

As discussed in the Introduction, whereas the chiral structure of the Standard Model prevents it
from accommodating any more than $N=1$ supersymmetry, any extension of the Standard Model at
the TeV scale would contain vector-like fermions, and hence
could accommodate $N=2$ supersymmetry. A first window on this doubling up of supersymmetry
could be provided by the Higgs sector. The two Higgs supermultiplets of the MSSM form a vector-like
pair, and thus could accommodate $N=2$ supersymmetry. Measurements of the $h(125)$ boson
and searches for heavier Higgs bosons in LHC Run~1 can already be used to probe this possibility.

In order to analyze this option, we have introduced an $h$2MSSM scenario in which the stop sector
is assumed to lift the $h$ mass from its tree-level value to the measured $m_h = 125$~GeV through
one-loop radiative corrections. This scenario is exactly analogous to the $h$MSSM scenario proposed
previously within the usual $N=1$ MSSM context \cite{hMSSM}. An interesting aspect of the $h$2MSSM scenario is
that much smaller stop masses are required to obtain $m_h = 125$~GeV than are needed in the $h$MSSM,
for any given values of $m_A$ and $\tan \beta$.

Another interesting feature of the $N=2$ extension of the MSSM is that the heavy Higgs bosons $H, A, H^\pm$
decouple from the massive vector bosons $W^\pm, Z^0$ at the tree level. This observation is subject
to radiative corrections, but the decoupling limit is a sufficiently good approximation that current searches
for $H \to W^+ W^-$, $Z^0 Z^0$ and $A \to Z h$ do not constrain the $h$2MSSM significantly. On the other hand,
the constraints from the decays of the heavy Higgs bosons to fermions are the same in the $h$2MSSM
as in the $h$MSSM.

The most stringent constraints on the $h$2MSSM come from LHC Run~1 measurements of the $h(125)$
couplings, including those to fermions, massive and massless gauge bosons.
However, these constraints are considerably weaker than in the $h$MSSM. We find that 
$m_A \gtrsim 185$~GeV is possible in the $h$2MSSM, whereas $m_A \gtrsim 350$~GeV is required
in the $h$MSSM. 

Looking to the future, we have also calculated the possible $N=2$ Higgs sector contributions
to anomalous couplings of the $h(125)$ boson. Current limits on these couplings do not constrain
the $N=2$ model, but this may be an interesting window for future measurements at the LHC and
elsewhere.

Doubling up supersymmetry opens up the possibility that supersymmetric Higgs bosons and stop squarks
could be significantly lighter than in the MSSM. Maybe Run~2 of the LHC will discover not just one 
supersymmetry, but two?

\section*{Acknowledgements:} 
The work of JE and JQ is supported partly by STFC Grant ST/L000326/1, and that of VS by STFC Grant ST/J000477/1.  
JE thanks the CERN Theoretical Physics Department for its hospitality, and VS thanks Martin Gorbahn for useful discussions.

\begin{small}

\end{small}

\end{document}